\documentclass[12pt]{article}
\usepackage[a4paper]{geometry}
\usepackage{jcapmod}
\usepackage{amssymb}
\usepackage{bm}
\usepackage{latexsym}
\usepackage{amssymb}
\usepackage{amsmath}
\usepackage{amsfonts}
\usepackage{colortbl}
\usepackage{multirow}
\usepackage{array}
\usepackage{booktabs}
\usepackage{rotating}
\usepackage{graphicx}

\def\clap#1{\hbox to 0pt{\hss#1\hss}}

\def\mathclap{\mathpalette\mathclapinternal}

\def\mathclapinternal#1#2{\clap{$\mathsurround=0pt#1{#2}$}}

\renewcommand{\geq}{\geqslant}
\renewcommand{\leq}{\leqslant}

\newcommand{\vect}[1]{\bm{\mathrm{{#1}}}}

\newcommand{\Mp}{M_{\mathrm{P}}}

\newcommand{\etal}{et al.}

\newcommand{\Ps}{\mathcal{P}}
\newcommand{\fNL}{f_{\mathrm{NL}}}
\newcommand{\tauNL}{\tau_{\mathrm{NL}}}
\newcommand{\gNL}{g_{\mathrm{NL}}}

\newcommand{\cs}{c_s}

\newcommand{\iprod}[2]{\langle{{#1}},{{#2}}\rangle}

\newcommand{\Rbasis}[1]{\mathcal{R}'_{{#1}}}
\newcommand{\Qbasis}[3]{\mathcal{Q}_{({#1},{#2},{#3})}}
\newcommand{\qpoly}[1]{q_{{#1}}}

\newcommand{\Local}{\textsf{local}}
\newcommand{\Equilateral}{\textsf{equilateral}}
\newcommand{\Orthogonal}{\textsf{orthogonal}}
\newcommand{\Enfolded}{\textsf{enfolded}}

\DeclareMathOperator{\Or}{O}

\DeclareMathOperator{\sgn}{sgn}

\newcommand{\transpose}{\mathsf{T}}

\renewcommand{\d}{\mathrm{d}}

\newcommand{\para}[1]{\par\vspace{2mm}\noindent\emph{{#1}}.---}

\newcommand{\tmark}[1]{\ensuremath{^{{#1}}}}

\newcolumntype{Q}{>{$\displaystyle}l<{$}}
\newcolumntype{q}{>{\columncolor[gray]{0.9}$\displaystyle}l<{$}}
\newcolumntype{R}{>{$\displaystyle}r<{$}}
\newcolumntype{S}{>{$\displaystyle}c<{$}}
\newcolumntype{s}{>{\columncolor[gray]{0.9}$\displaystyle}c<{$}}
\newcolumntype{T}{>{\columncolor[gray]{0.9}}c<{}}

\newsavebox{\tableA}
\newsavebox{\tableB}
\newlength{\tblw}

\newsavebox{\boxplot}
\newsavebox{\boxplota}
\newlength{\plotw}
\newlength{\plotwa}

\newcommand{\tablepreamble}{\small%
	\heavyrulewidth=.08em%
	\lightrulewidth=.05em%
	\cmidrulewidth=.03em%
	\belowrulesep=.65ex%
	\belowbottomsep=0pt%
	\aboverulesep=.4ex%
	\abovetopsep=0pt%
	\cmidrulesep=\doublerulesep%
	\cmidrulekern=.5em%
	\defaultaddspace=.5em%
	\renewcommand{\arraystretch}{1.6}}

\notoc

\begin{document}

	\title{Decoding the bispectrum of single-field inflation}

	\author[a]{Raquel H. Ribeiro,}
	\author[b]{David Seery}

	\affiliation[a]{Department of Applied Mathematics and Theoretical Physics\\
	Centre for Mathematical Sciences, Wilberforce Road \\
	Cambridge CB3 0WA, United Kingdom} 

	\affiliation[b]{Astronomy Centre, University of Sussex \\
	Falmer, Brighton BN1 9QH, United Kingdom}

	\emailAdd{R.Ribeiro@damtp.cam.ac.uk}
	\emailAdd{D.Seery@sussex.ac.uk}

	\abstract{Galileon fields 
	arise naturally from the decoupling limit
	of massive gravities,
	and possess special self-interactions which are protected
	by a spacetime generalization of
	Galilean symmetry.
	We briefly revisit the inflationary phenomenology of Galileon
	theories.
	Working from recent computations of the fluctuation
	Lagrangian to cubic order
	in the most general model with second-order equations of motion,
	we show that a distinct shape is present
	but with suppressed amplitude.
	A similar
	shape has been found in other higher-derivative models.
	It may be visible in a theory tuned to suppress the
	leading-order shapes, or if the overall bispectrum has
	large amplitude.
	Using a partial-wave expansion of the bispectrum,
	we suggest a possible origin for the frequent
	appearance of this shape.
	It follows that models with very disparate microphysics
	can produce very similar bispectra.
	We argue that it may be more profitable to distinguish
	these models by searching for relations between the
	amplitudes of these common shapes.
	We illustrate this method using the examples of
	DBI and $k$-inflation.}
	

\vspace*{2cm}

	\maketitle

	\section{Introduction}
	\label{sec:introduction}
	
	Over the last few decades, advances in observational
	cosmology have led to a
	detailed picture of the microwave sky
	\cite{Komatsu:2010fb, Larson:2010gs, Jarosik:2010iu},
	now known to be almost smooth with fluctuations at the level of
	1 part in $10^5$.
	Among the most popular proposals for the mechanism which seeded
	these small perturbations
	is inflation,
	in which the universe underwent 
	a quasi-de Sitter expansion \cite{Guth:1980zm, Guth:2005zr}. 
	When combined with quantum mechanics, 
	inflation allows the growth of
	density fluctuations which
	classicalize after horizon crossing \cite{Bardeen:1983qw}.
	They are subsequently
	imprinted in the CMB as temperature anisotropies. 
	The statistics of the observable temperature field
	map directly from the
   	primordial density perturbation,
   	which in turn depends on the microphysics
   	governing
	the very early universe.

	The link with microscopic physics suggests that it may be
	possible to distinguish different models giving rise to
	inflation by studying three- and higher $n$-point correlations
	\cite{Komatsu:2009kd}.
 	Current observations suggest that departures from 
 	gaussianity
 	are small, but
	non-gaussian correlations are generated
	at a low level by most
	microscopic models
	and it remains worthwhile to search for them.
 	Computationally and observationally the 
 	best place to look
 	is the bispectrum,
 	which contains multiple sources of information:
 	a number of distinct shapes \cite{Babich:2004gb}
 	or ``channels''---%
 	analogous to, but more complicated
 	than, the Mandelstam channels of $2 \rightarrow 2'$
 	scattering---%
 	together with their amplitudes.
	The shapes depend on the three-body interactions responsible for
	generating nontrivial correlations, and
	the amplitudes measure their relative
	importance.
	For reviews, see
	Refs.~\cite{Chen:2010xka,Koyama:2010xj}.
	Recent work employing the bispectrum as a discriminant of
	microphysics includes Refs.
	\cite{Burrage:2010cu,Baumann:2011su,Creminelli:2010qf,Gao:2011qe,
	DeFelice:2011uc,RenauxPetel:2011sb,RenauxPetel:2011uk}.
	
	If gravity is modified in the infrared, perhaps in a way which 
	accounts for our presently accelerating phase,
	then this may leave traces in the primordial density fluctuation
	\cite{Vazquez:2008wb,Gao:2010um}.
	Recently there has been interest in ``Galileon'' fields,
	which can be thought of as an effective \emph{short}-distance
	description of
	longitudinal graviton modes
	near the decoupling limit
	of massive gravity
	\cite{deRham:2010ik,deRham:2010kj,Hassan:2011hr},
	where $\Mp \rightarrow \infty$
	while the cutoff remains fixed.
	A clear discussion is given in the review by
	Hinterbichler~\cite{Hinterbichler:2011tt}.
	
	A Galileon singlet owes its name 
	to invariance under the transformation
	\begin{equation}
		\phi(x) \rightarrow  \phi(x) + b_{\mu}x^{\mu} + c ,
		\label{eq:galileon-xfm}
	\end{equation}
	for constant $b_\mu$ and $c$.
	Eq.~\eqref{eq:galileon-xfm} is a spacetime version of
	a Galilean transformation,
	first noticed in the DGP model
	\cite{Dvali:2000hr,Deffayet:2001pu}.
	It
	incorporates the shift symmetry $\phi \rightarrow \phi + c$,
	which implies that
	$\phi$ can support
	a long-lived inflationary epoch in the early universe.
	Indeed, in this scenario the principal difficulty is \emph{ending}
	inflation.
	To do so one must stabilize the field,
	typically by introducing a potential.
	Because this breaks~\eqref{eq:galileon-xfm} by design,
	further Galilean-violating terms may be generated radiatively.
	It may then be technically unnatural to
	start from an
	action which approximately respects~\eqref{eq:galileon-xfm}.

	In Ref.~\cite{Burrage:2010cu} it was argued that this difficulty
	can be avoided.
	Taking the potential to be sufficiently mild, Galilean-violating
	radiative corrections are suppressed,
	making a Lagrangian dominated by
	terms respecting~\eqref{eq:galileon-xfm} technically natural.
	The prospects for inflation have been studied by
	several authors, 
	often relaxing invariance under~\eqref{eq:galileon-xfm}
	and requiring
	only the weaker condition of second-order equations of
	motion
	\cite{Kobayashi:2010cm, Mizuno:2010ag, Kobayashi:2011pc,
	RenauxPetel:2011dv,
	Kobayashi:2011nu, Gao:2011qe}.
	The most general action of this type was
	written down over thirty-five years ago
	by Horndeski~\cite{Horndeski:1974},
	and later revisited by several authors 
	\cite{Charmousis:2011bf, Kobayashi:2011nu, Deffayet:2009mn,Deffayet:2011gz}.
	The first bispectrum estimate
	was obtained by
	Mizuno \& Koyama \cite{Mizuno:2010ag}, who worked with a model
	where the most relevant Lagrangian operator was $(\partial \phi)^2
	\Box \phi$. The result for the complete covariant Galileon,
	in the decoupling limit, was given in
	Ref.~\cite{Burrage:2010cu}.
	A class of related
	of models was considered by Creminelli {\etal}
	\cite{Creminelli:2010qf}.
	More recently, Gao \& Steer~\cite{Gao:2011qe}
	(see also Renaux-Petel~\cite{RenauxPetel:2011sb})
	and de Felice \& Tsujikawa~\cite{DeFelice:2011uc}
	obtained the bispectrum for the entire
	Horndeski action and retained the coupling to gravity.

	In simple models, the bispectrum is practically determined
	by Lorentz invariance of the underlying Lagrangian
	and the unbroken spatial symmetries of
	de Sitter space \cite{Cheung:2007st}.
	In Galileon models some of this simplicity is lost,
	and the bispectrum can be more complicated.
	Nevertheless, Creminelli {\etal} were able to conclude that
	no Lagrangian operators became available beyond those
	which could already be realized in simpler models
	\cite{Creminelli:2010qf}. Therefore
	the distinctiveness of the Galileon bispectrum lies
	only in their relative amplitudes. In practice this means that the
	models could be difficult to distinguish.
	The recent analyses of
	Refs. \cite{Gao:2011qe, DeFelice:2011uc, RenauxPetel:2011sb}
	have extended this disappointing
	conclusion to the full Horndeski
	action.
	
	Although no new operators are present, the number of linearly
	independent \emph{shapes} depends on the number of arbitrarily
	adjustable coefficients in the Lagrangian.
	In this paper we revisit the question of how many shapes
	should be expected.
	At leading order, we
	find one extra channel
	typically becomes available---although with suppressed amplitude---%
	which is similar
	to the shape identified by Creminelli {\etal}~\cite{Creminelli:2010qf}
	and rediscovered at next-order in $P(X, \phi)$ models
	in Ref.~\cite{Burrage:2011hd}.%
		\footnote{It was remarked in Ref.~\cite{Burrage:2011hd}
		that these shapes are visually quite similar.
		They have a relatively strong cosine~\cite{Babich:2004gb},
		typically of order $\sim 0.9$.
		However,
		there are differences which we will discuss
		in \S\ref{sec:partial-wave}.}
	This apparent universality is surprising;
	although the action used by Creminelli {\etal}
	is ``Galileon,'' it is not closely related to that of
	Refs.~\cite{Burrage:2010cu, Burrage:2011hd}.
	Therefore the similarity of their bispectra
	could not easily have been anticipated:
	they are intricate objects having no
	simple connexion to each other.
	We employ a partial-wave expansion of the bispectrum
	to explain some features of this shape.
	We find that the basis suggested by
	Fergusson {\etal}~\cite{Fergusson:2009nv}
	is useful in describing the primordial bispectrum,
	and gives guidance concerning the distinguishable shapes which
	can be expected.
	We give a brief sketch of how a decomposition 
	into these partial waves can be used to derive
	``consistency equations,''
	which express predictions of the theory as
	relations between observable quantities.
	By determining whether these relations are satisfied,
	it is possible to rule out classes of scenarios.
	
	\para{Outline}%
	This paper is organized as follows.
	In~\S\ref{sec:galileon_inflation} we briefly review 
	Galileon inflation and explore the bispectrum shapes
	at leading order in slow-roll. We show there is an 
	orthogonal shape with suppressed amplitude, which
	turns out to be related to one
	present in other single-field models.
	To understand the recurrence of this shape, 
	in \S\ref{sec:partial-wave} we apply a decomposition 
	of bispectrum shapes in terms of an orthogonal basis. We 
	argue that
	it may be possible to derive
	model independent tests using the 
	coefficients of these linear decompositions as appropriate observables.
	We conclude in~\S\ref{sec:conclusions}. 

	We work in units where $\hbar = c = 1$, and
	define 
	the reduced Planck mass to be
	$\Mp = (8\pi G)^{-1/2}$, where $G$ is Newton's gravitational constant.
	When discussing the common bispectrum templates, we denote
	them
	``{\Equilateral},'', ``{\Orthogonal},''
	``{\Enfolded},'' and ``{\Local}''
	to distinguish the {\Orthogonal} template and other shapes which may
	or may not be orthogonal to each other.

	\section{Shapes in single-field inflation}
	\label{sec:galileon_inflation}
	
	\paragraph{Background.}
	Beyond the DGP model,
	the first Galileon theories were constructed by
	Nicolis {\etal}~\cite{Nicolis:2008in},
	who restricted their discussion to a Minkowski background.
	Their theory was designed to produce second-order
	equations of motion, even though the action included
	high-order combinations of derivatives.
	Higher-order equations of motion
	would have implied propagating ghosts,
	and a loss of unitarity when interpreted as a quantum theory.
	The success of Nicolis {\etal} in achieving second-order
	equations of motion was later understood from a more general point of
	view~\cite{deRham:2010eu}.
	
	For application to the early universe, the Galileon must be promoted
	to curved spacetime.
	To protect the important property
	of second-order equations of motion, one must introduce non-minimal
	couplings to the curvature. The result is the
	``covariant'' theory of Deffayet {\etal}~\cite{Deffayet:2009wt}.
	Later work on curved backgrounds includes
	Refs.~\cite{Deffayet:2011gz,Goon:2011qf,Goon:2011uw,Trodden:2011xh,
	Burrage:2011bt}.
	We write the Galileon field $\phi$.
	On a de Sitter background, where $a(t)=\exp (Ht)$,
	it is spatially homogeneous
	and depends only on time, $t$.
	The action is
	\begin{equation}
		S = \int \d^4 x \; a^3 \bigg\{ 
			\frac{c_2}{2} \dot{\phi}^2
			+ \frac{2c_3 H}{\Lambda^3} \dot{\phi}^3
			+\frac{9c_4H^2}{2\Lambda^6} \dot{\phi}^4
			+\frac{6c_5H^3}{\Lambda^9} \dot{\phi}^5
			-V(\phi)
		\bigg\} \; .
	\label{eq:ds-action}
	\end{equation}
	The potential $V(\phi)$ is chosen to softly break the Galilean invariance
	and is necessary to end inflation, as discussed
	in~\S\ref{sec:introduction}.
	The scale $\Lambda$ is the na\"{\i}ve cutoff of the theory.
	In practice, a Vainshtein effect can allow~\eqref{eq:ds-action}
	to describe fluctuations at higher energies~\cite{Vainshtein:1972sx}.
	The most general models allow
	the $c_i$ to be unconstrained, unless one demands compatibility with
	late-time cosmological or laboratory tests
	\cite{Burrage:2010rs,Gannouji:2010au,Ali:2010gr,Brax:2011sv}.
	This is optional because it need not be supposed that $\phi$ is
	active in the post-inflationary universe. 
	If the Galileon field was present only during 
	inflation,
	then 
	constraints on $c_i$ follow by demanding agreement 
	with the standard inflationary observables.
	Where the Galileon theory arises from the decoupling limit of
	a ghost-free massive gravity, other constraints may arise
	\cite{deRham:2010ik}.

	\paragraph{Fluctuations.}
	We briefly review the calculation of inflationary perturbations.
	The Horndeski action is sufficiently general to include
	the covariant Galileon together with other theories which
	do not exhibit Galilean symmetry
	\cite{Deffayet:2010qz,Kobayashi:2010cm,Kobayashi:2011pc,
	Deffayet:2011gz,Kobayashi:2011nu}.
	It
	turns out to be
	no more complicated to give the analysis
	for the Horndeski action, which we do for the sake of
	generality.
	Including gravitational effects,
	three-body interactions among
	scalar fluctuations in Horndeski's model
	are described by the action~\cite{Gao:2011qe,DeFelice:2011uc,
	RenauxPetel:2011sb}
	\begin{equation}
		\begin{split}
		S \supseteq 
		\int \d^3 x \, \d\tau \; &\Big\{ 
			a^2 M^2 \left[ \zeta'^2 - \cs^2 (\partial \zeta)^2
			\right]
			+ a \Lambda_1 \zeta'^3 +a^2 \Lambda_2 \zeta \zeta'^2
			+ a^2 \Lambda_3 \zeta (\partial \zeta)^2 \\
			&
			+ a^2 \Lambda_4 \zeta' \partial_i \zeta
				\partial^i (\partial^{-2}\zeta')
			+ a^2 \Lambda_5 \partial^2 \zeta
				(\partial_i \partial^{-2}\zeta')^2	
		\Big\} .
		\end{split}
		\label{eq:zeta-action}
	\end{equation}
	In writing this action we have exploited our freedom to
	integrate by parts, and
	removed redundant
	couplings using the equations of motion.
	Primed quantities are differentiated with respect to conformal time,
	$\tau = \int_\infty^t \d t / a(t)$.
	The field $\zeta$ is the primordial curvature perturbation,
	and is related to the field fluctuation at linear order
	by the usual rule
	$\zeta = H \delta\phi / \dot{\phi}$.
%
	Its fluctuations propagate at the phase velocity $\cs$.
	The mass $M$ sets the scale of the action.
	Specializing to
	the covariant Galileon would correspond to specific assignments
	of the $\Lambda_i$, but
	the detailed form of these coefficients will not be important
	for our discussion.
	For Horndeski's general action the $\Lambda_i$ can be adjusted
	independently.

	\subsection{Shapes}
	\label{sec:shapes}
	\label{sec:new_shape}

	\paragraph{Inner product.}
	Conservation of 3-momentum in the bispectrum
	requires that the momenta $\vect{k}_i$ generate a triangle in
	momentum space.
	The bispectrum is a function on this space of triangles.
	Babich \etal~\cite{Babich:2004gb} described its
	functional form as
	the ``shape'' of the bispectrum and introduced a measure
	to distinguish qualitatively different shapes.
	Define an inner product between two
	bispectra $B_1$, $B_2$ by the rule
	\begin{equation}
		\iprod{B_1}{B_2}
		\equiv \;
			\int\limits_{\mathclap{\text{triangles}}}
			\d k_1 \, \d k_2 \, \d k_3 \;
			S_1(k_1, k_2, k_3) S_2(k_1, k_2, k_3)
			,
 		\label{eq:inner-prod-def}
	\end{equation}
	where $B_i = (k_1 k_2 k_3)^{-2} S_i$, and $S_i$ is called
	the \emph{shape}.
	The norm of any bispectrum is $\|B\| = \iprod{B}{B}^{1/2}$,
	and
	the cosine between two bispectra
	is the normalized inner product,
	$\cos (B_1, B_2) \equiv \iprod{B_1}{B_2} / \|B_1\| \|B_2\|$.
	Further details can be obtained from
	Refs.~\cite{Babich:2004gb,Fergusson:2008ra,Fergusson:2009nv}.
	Our conventions, particularly for assigning meaning
	to divergences in the squeezed limit, follow~Ref.~\cite{Burrage:2011hd}.

	The bispectrum, $B$, is defined to satisfy
	\begin{equation}
		\langle \zeta(\vect{k}_1) \zeta(\vect{k}_2) \zeta(\vect{k}_3)
		\rangle
		= (2\pi)^3 \delta(\vect{k}_1 + \vect{k}_2 + \vect{k}_3)
		B(k_1, k_2, k_3) .
	\end{equation}
	For a general Horndeski model, $B$
	will receive contributions at leading order from all operators
	in~\eqref{eq:zeta-action}.
	This yields $B = (k_1 k_2 k_3)^{-2} \sum_a S_a$,
	where each operator yields a shape $S_a$, and $a$ labels
	the distinct operators in the Lagrangian.
	We plot
	the $S_a$ in table \ref{table:shapes-leading-order},
	computed at leading order in the slow-roll approximation,
	and quote their cosines with the common templates 
	of CMB analysis in table~\ref{table:leading_overlaps}.
	The $\zeta'^3$, $\zeta'\partial \zeta \partial \partial^{-2} \zeta'$
	and $\partial^2 \zeta (\partial \partial^{-2} \zeta')^2$ shapes
	are
	highly correlated with the {\Equilateral} template.
	The $\zeta \zeta'^2$ and $\zeta (\partial \zeta)^2$
	shapes are correlated with the {\Local} template.
	In most cases there is a moderate overlap with
	the {\Enfolded} template.
	In some cases, corrections at subleading order (``next-order'')
	in the slow-roll expansion may become important.
	These have been catalogued in Ref.~\cite{Burrage:2011hd},
	to which we refer for details,
	for the 
	action~\eqref{eq:zeta-action} with arbitrary $\Lambda_i$.
	These corrections therefore apply to
	an arbitrary action of Horndeski type.

	\paragraph{Bispectrum.}	
	Factoring out an overall normalization,
	the shape $S$ of the bispectrum can be written
	\begin{equation}
		S \propto
		\mbox{} 
		\alpha  S_{\zeta'^3} +
		\beta   S_{\zeta \zeta'^2} + \gamma S_{\zeta (\partial\zeta)^2}+
		\delta   S_{\zeta' \partial_i \zeta \partial^i (\partial^{-2}\zeta')}+
		\omega   S_{\partial^2 \zeta (\partial_i \partial^{-2}\zeta')^2} ,
		\label{eq:start_ortho}
	\end{equation}
	where $\alpha, \beta, \gamma, \delta, \omega$ are 
	rescaled versions of the coefficients $\Lambda_i$.
	In a generic model we could perhaps expect all these ratios to be
	order unity, although in specific cases some may be much smaller.
	By adjusting these coefficients it is possible to find a
	``critical surface'' on which $B$ becomes orthogonal to
	some specified set of templates.
	To be concrete, we choose the
	set $Z = \{ \text{\Equilateral}, 
	\text{\Local}, 
	\text{\Enfolded} \}$.
	The bispectrum can be written
	\begin{equation}
		S \propto
		\left(
			\begin{array}{ccc}
				\delta & \omega 
			\end{array}
		\right)
		\left(
			\begin{array}{c}
				S_{\delta} \\
				S_{\omega} 
			\end{array}
		\right)
		+
		\left(
			\begin{array}{ccc}
				a & b & c
			\end{array}
		\right)
		\left(
			\begin{array}{c}
				S_{\zeta'^3} \\
				S_{\zeta \zeta'^2} \\
				S_{\zeta (\partial \zeta)^2}
			\end{array}
		\right)
		.
	\end{equation}
	Here, the new shapes $S_\delta$ and $S_\omega$ are
	orthogonal by construction to
	each template in $Z$.
	The coefficients $\delta$ and $\omega$
	act as coordinates on the
	subspace of bispectra which are
	also orthogonal to these templates.
	Likewise, $a$, $b$ and $c$ act as coordinates labelling
	departures from this critical subspace.
	They are defined by
	\begin{subequations}
		\begin{align}
			\alpha 
			& \approx 2.394 \delta + 2.208 \omega + a\\ 
			\beta & \approx 0.473 \delta + 0.642 \omega + b\\
			\gamma  & \approx -0.183 \delta - 0.248 \omega	+ c.
			\label{eq:alphabetagamma}
		\end{align}
	\end{subequations}
	The shapes $S_\delta$ and $S_\omega$ satisfy
	\begin{subequations}
		\begin{align}
			S_{\delta} & \approx
				2.394 S_{\zeta'^3}
				+ 0.473S_{\zeta \zeta'^2}
 				- 0.183 S_{\zeta (\partial \zeta)^2}
 				+ S_{\zeta' \partial_i \zeta \partial^i (\partial^{-2}\zeta')}
 			\\
 			S_{\omega} & \approx
 				2.208 S_{\zeta'^3}
				+ 0.642 S_{\zeta \zeta'^2}
 				- 0.248 S_{\zeta (\partial \zeta)^2}
 				+ S_{\partial^2 \zeta (\partial_i \partial^{-2}\zeta')^2}
 			.
 		\end{align}
	\end{subequations}
	Although we did not require it,
	these shapes
	are also highly orthogonal to the
	``{\Orthogonal}'' template introduced by Senatore
	{\etal}~\cite{Senatore:2009gt}.
	(See also~\S\ref{sec:partial-wave}.)
	But they need not be orthogonal amongst themselves. To measure
	\emph{independent} combinations from data typically requires
	a dedicated template which has negligible overlap with other
	combinations. We follow the procedure of
	Refs.~\cite{Senatore:2009gt,Chen:2010xka}.
	The inner product matrix is $C_{ij} \equiv S_i \cdot S_j$.
	It is diagonalized by an orthogonal matrix
	$\vect{P}$ whose columns are formed from the eigenvectors
	of $\vect{C}$. Setting $a = b = c =  0$
	and writing
	$\vect{x} = ( \begin{array}{ccc} \delta  & \omega  \end{array} )$,
	$\vect{S} = ( \begin{array}{ccc} S_{\delta} & S_{\omega}
	\end{array} )^\transpose$,
	the part of bispectrum on the
	critical subspace can be written
	$B^\parallel \propto \vect{q}\vect{H}$,
	where $\vect{q} \equiv \vect{x} \vect{P}$
	and $\vect{H} \equiv \vect{P}^\transpose \vect{S}$.

	The shapes $S_{\zeta \zeta'^2}$ and $S_{\zeta (\partial \zeta)^2}$
	have {\Local}-type divergences,
	which can be subtracted by
	taking a suitable linear combination.
	This leaves four
	independent terms, from which we wish to construct a linear combination
	orthogonal to three templates. We should expect a unique solution.
	This can be extracted from $\vect{H}$, and is
	\begin{equation}
		S_H = -0.805 S_{\delta} +  0.593 S_{\omega} .
		\label{eq:o-def}
	\end{equation}
	This procedure discards the independent linear combination
	of $S_{\zeta\zeta'^2}$ and $S_{\zeta (\partial \zeta)^2}$.
	For practical purposes, we expect
	its divergence in the squeezed limit
	to make it almost indistinguishable from the
	{\Local} template.
	We ignore it in the equations which follow, such as~\eqref{eq:b-generic},
	although in principle one should remember that it is present.
	In table~\ref{table:ortho-approx} we plot $S_H$
	together with the ``orthogonal'' shapes
	which were encountered by Creminelli {\etal}~\cite{Creminelli:2010qf}
	and Ref.~\cite{Burrage:2011hd}.

	These shapes are all similar.
	When plotted using the method of Babich {\etal}~\cite{Babich:2004gb}
	the shape has a wavelike appearance.
	In the Fergusson {\etal}~\cite{Fergusson:2008ra}
	plots of table~\ref{table:ortho-approx},
	they smoothly converge to zero in the squeezed limit but
	exhibit distinctive ``teardrop''
	or drumlin-shaped features near the corners of the triangle.
	The $S_H$-shape of Eq.~\eqref{eq:o-def} is closer to the
	shape of Creminelli~{\etal} than the $P(X,\phi)$-based shape
	of Ref.~\cite{Burrage:2011hd}.
	However, the overall similarity suggests 
	there is little difference in available \emph{shapes} between
	different microphysical models. We will return to this issue
	in~\S\ref{sec:partial-wave}.
	
	The shape $S_H$ will occur in a typical bispectrum with
	coefficients which depend on $\omega$ and $\delta$.
	We find
	\begin{equation}
		S = (0.593 \omega - 0.805\delta) S_H
		+ \left(
			\begin{array}{c}
				\alpha - 2.394 \delta - 2.208 \omega  \\
				\beta -0.473 \delta - 0.642  \omega \\
				\gamma +0.183 \delta + 0.248  \omega
			\end{array}
		\right)
		\left(
			\begin{array}{c}
				S_{\zeta'^3} \\
				S_{\zeta \zeta'^2} \\
				S_{\zeta (\partial \zeta)^2}
			\end{array}
		\right) .
		\label{eq:b-generic}
	\end{equation}
	How significant is its contribution?
	Since all prefactors will generically be of order unity,
	the question reduces to the relative magnitudes of $S_H$ and
	the $S_a$. We find $\| S_H \| \approx 10^{-2}$,
	whereas $\| S_{\zeta'^3} \| \approx 1$.
	The precise values assigned to $\| S_{\zeta \zeta'^2} \|$
	and $\| S_{\zeta (\partial \zeta)^2} \|$
	depend how their squeezed divergences are regulated,
	and therefore do not form a fair basis for comparison.
	Cutting out the divergent regions one finds
	$\| S_{\zeta \zeta'^2} \|$
	and $\| S_{\zeta (\partial \zeta)^2} \|$
	to be of order $10^{1}$ to $10^{2}$.
	We conclude that $S_H$
	has an amplitude suppressed by roughly $10^3$
	to $10^{4}$ compared with
	the leading-order shapes.
	All of these are well-matched by the standard templates.
	For the new shape $S_H$ to be visible requires \emph{either}
	\begin{itemize}
		\item The leading order shapes to be suppressed,
		so that $a \approx b \approx c \approx 0$ to an accuracy
		of about a few parts in $10^3$ to $10^4$.
		This could happen in a specific model,
		but requires some tuning.
		\item The overall amplitude of the bispectrum
		to be sufficiently large that the suppressed
		$S_H$ shape is visible.
		Without a dedicated analysis of the signal-to-noise
		available in the $S_H$-channel for a CMB survey,
		it is not possible to know how large the bispectrum must be.
		However, it is unlikely that the signal to noise for $S_H$
		will be dramatically better than that for the {\Equilateral} template.
		Therefore, it seems reasonable to suggest that
		the leading-order operators would have to produce
		$|\fNL^{\mathrm{eq}}| \gtrsim 100$
		in order for the $S_H$-shape to be visible.
		This is on the boundary of present-day experimental
		sensitivity
		\cite{Creminelli:2005hu,Creminelli:2006rz,
		Senatore:2009gt,Komatsu:2010fb}.
	\end{itemize}

	\section{Partial-wave decomposition of the bispectrum}
	\label{sec:partial-wave}
	
	It is natural to ask why the various
	shapes obtained in~\S\ref{sec:new_shape}
	and Refs.~\cite{Creminelli:2010qf,Burrage:2011hd}
	are so similar.
	One answer is that they have all been constructed by taking
	linear combinations of similar-looking bispectra
	in a way designed to produce shapes orthogonal to the standard
	templates. Since the inputs are similar, so are the outputs.
	
	Although this answer is correct, it does not make clear why a linear
	combination of dome-shaped bispectra should produce the
	characteristic drumlin shapes of table~\ref{table:ortho-approx}.
	The drumlin increases the number of nodes or anti-nodes
	exhibited by the bispectrum.
	One can think of its emergence
	in a similar way to taking two
	almost pure Fourier harmonics
	and constructing an orthogonal function.
	The result will be approximately
	the next available Fourier harmonic.
	Therefore, to obtain a more quantitative description
	one is led to decompose the bispectrum into some analogue of
	Fourier modes. The underlying triangular geometry
	is different to the flat intervals which yield Fourier harmonics,
	so the appropriate analogue will be a generalized partial wave.
	
	\paragraph{Harmonic decomposition.}
	Partial-wave decompositions have been usefully applied to
	correlation functions, in the form of
	scattering amplitudes, since the early days of quantum field theory.
	In $WW$ scattering, partial-wave methods give guidance concerning
	the energy scale where
	the Standard Model without a Higgs boson loses perturbative
	unitarity.
	Similar ideas underlie, for example, the method of complex angular momenta
	and Regge theory.
	They have not been widely applied to
	inflationary correlation functions, although
	Fergusson {\etal} \cite{Fergusson:2008ra,Fergusson:2009nv}
	introduced a number of partial-wave decompositions
	and emphasized their computational efficiency.
	We largely follow their method and notation.%
		\footnote{Physical conclusions must be independent of the
		basis, but the analysis may be made simpler by an
		appropriate choice.
		For comparison with the Fergusson~{\etal}
		basis, we have repeated the analysis using
		Bessel functions \cite{Fergusson:2008ra}.
		With this choice, convergence is much slower.
		A different decomposition was used by
		Meerburg~\cite{Meerburg:2010ks}.}
	
	Fergusson {\etal} suggested
	writing each shape function in the form
	\begin{equation}
		S(k_1, k_2, k_3)
		= \sum_{n} \alpha_n \Rbasis{n}(k_1, k_2, k_3) ,
		\label{eq:decompositionR}
	\end{equation}
	for some coefficients $\alpha_n$ and a set
	of dimensionless
	basis functions $\Rbasis{n}$ which are orthonormal in the
	inner product~\eqref{eq:inner-prod-def}.%
		\footnote{The functions we are denoting $\Rbasis{n}$
		are only a subset of those constructed by
		Fergusson~{\etal}~\cite{Fergusson:2008ra,Fergusson:2009nv}
		and labelled $\mathcal{R}_n$.
		The $\Rbasis{n}$ form a basis on a fixed slice
		at constant $k_t = k_1 + k_2 + k_3$.
		They are suitable for expansion of an approximately
		scale-invariant primordial bispectrum.
		The Fergusson~{\etal} $\mathcal{R}_n$
		are not scale-invariant and are orthonormal in
		a three-dimensional inner product which accounts for variation in
		$k_t$.
		Our $\Rbasis{n}$ are constructed using precisely the same
		procedure as the $\mathcal{R}_n$,
		but because many of the $\Rbasis{n}$ are degenerate
		purely as a function of shape (but not scale)
		they are projected out of the $\Rbasis{n}$.
		It is in this sense that
		the $\Rbasis{n}$ form a sparse
		subset of the $\mathcal{R}_n$.}
	The choice of $\Rbasis{n}$ was motivated by numerical
	considerations, as follows.
	Define a complete set of orthonormal polynomials $\qpoly{p}(x)$
	on the unit
	interval $x \in [0,1]$ with measure $w(x)$
	and introduce quantities $\Qbasis{p}{q}{r}$ satisfying
	\begin{equation}
		\Qbasis{p}{q}{r} =
			\qpoly{p}(2 k_1/k_t) \qpoly{q}(2 k_2/k_t) \qpoly{r}(2 k_3/k_t)
			+ \text{5 perms}.
	\end{equation}
	Fergusson~{\etal} chose $w$ to cancel
	an unwanted growth in the bispectrum at large $k$; for all details
	and the construction of the $q_p(x)$
	we refer to the original
	literature~\cite{Fergusson:2008ra,Fergusson:2009nv}.
	One may impose a fixed normalization for the
	$\Qbasis{p}{q}{r}$ if desired.
	They can be ordered by defining
	$\rho^2 = p^2 + q^2 + r^2$
	and sorting the $\Qbasis{p}{q}{r}$ in ascending order of $\rho$.
	Finally, the
	$\Rbasis{n}$ are constructed by Gram--Schmidt orthonormalization
	of the ordered $\Qbasis{p}{q}{r}$.
	It follows that
	the $\Rbasis{n}$ are a linear combination
	of separable functions.
	This leads to efficiencies
	in computation of CMB observables,
	which was the principal motivation for
	Refs.~\cite{Fergusson:2008ra,Fergusson:2009nv}.
	Because the $\Rbasis{n}$ are orthonormal,
	one can obtain the expansion coefficients $\alpha_n$
	for any bispectrum $B$
	using the inner product~\eqref{eq:inner-prod-def},
	\begin{equation}
		\alpha_n = \iprod{\Rbasis{n}}{B} .
		\label{eq:r-expansion}
	\end{equation}
	Note that $\| B \|^2 = \sum_n \alpha_n^2$, so one can
	interpret the ratio $\alpha_n^2 / \alpha_m^2$ as a measure of
	the relative
	importance of the $m^{\mathrm{th}}$ and $n^{\mathrm{th}}$
	modes.
	We plot the first few $\Rbasis{n}$ in Table~\ref{table:Rbasis}
	and quote $\alpha_n$ for the common templates in
	Table~\ref{table:Rtemplates}.
	The $n=0$ mode is a constant. The $n=1,2$ modes are a good match for
	the overall shape of both the
	{\Equilateral} and {\Orthogonal} templates.
	Strong features in the corners of the triangle,
	characteristic of the {\Local} shape,
	appear at higher $n$.
	
	\paragraph{Orthogonal combinations.}
	For our purposes, the usefulness of the $\Rbasis{n}$
	stems from the fact that
	the first three partial waves
	provide a very good description of the {\Equilateral}, {\Orthogonal}
	and {\Enfolded}
	templates.
	These can all be obtained by shifting the {\Equilateral} shape by
	a constant \cite{Senatore:2009gt,Chen:2010xka}.
	The $\Rbasis{0}$ shape is the constant shift.
	The ``first harmonic,''
	$\Rbasis{1}$, peaks in the equilateral limit,
	whereas $\Rbasis{2}$ peaks in the \emph{flattened} configuration,
	where $\alpha = \beta = 0$. (This makes the two smallest $k_i$
	equal to one-half of the largest $k_i$.)
	These two peaks accurately describe the characteristics
	of the {\Equilateral}/{\Orthogonal}/{\Enfolded} class
	\cite{Senatore:2009gt}.
	See also the discussion in Renaux-Petel~{\etal}~\cite{RenauxPetel:2011uk}.

	We quote expansion coefficients for the
	common templates in table~\ref{table:Rtemplates},
	obtained using Eq.~\eqref{eq:r-expansion}.
	For the reasons we have explained,
	the {\Equilateral}, {\Orthogonal} and {\Enfolded} templates
	are dominated by $\{ \Rbasis{0}, \Rbasis{1}, \Rbasis{2} \}$,
	with their coefficients diminishing for higher $n$.
	This explains why the shape $S_H$ of~\eqref{eq:o-def}
	has negligible overlap with the {\Orthogonal} template,
	even though this was not guaranteed by its construction.
	On the other hand,
	the {\Local} shape does not have a rapidly convergent expansion
	because its squeezed divergence requires a mixture of
	modes with $n \gg 1$.
	The net result is that the $\Rbasis{n}$-basis
	is reasonably well-adapted for an efficient description of
	the
	higher-derivative
	self-interactions of $\zeta$, which typically do not generate
	such divergences.
	
	One can regard the orthogonalization process described
	in~\S\ref{sec:new_shape}
	as suppressing the coefficients of
	$\{ \Rbasis{0}, \Rbasis{1}, \Rbasis{2} \}$.
	We give the expansion coefficients for the
	various ``new'' shapes
	in table~\ref{table:Rorthogonal}.
	Consulting these coefficients shows that the $\Rbasis{0}$
	shape is projected out entirely for
	the shape of Creminelli {\etal}
	and
	the $S_H$-shape
	of~\eqref{eq:o-def}.
	The situation for the $P(X, \phi)$ shape is more complicated,
	and requires a separate discussion. For the remainder of this
	section we exclude it from our analysis.
	For the other two shapes,
	the $n = 1,2$ harmonics are not completely removed but their
	amplitudes are significantly reduced.
	As with the analogous case of Fourier harmonics,
	the largest individual term in each orthogonalized shape
	is a nearby higher mode---in this case,
	the $n=3$ term. (This is the next highest, although recall that
	the precise
	ordering of the $\Rbasis{n}$ is somewhat arbitrary.)
	There is an admixture of the other harmonics with smaller
	amplitudes.
	Comparison with table~\ref{table:Rbasis}
	shows
	that the large $n=3$ contribution
	is essentially responsible
	for the common appearance of teardrops or drumlins.
	In practice, the broad hotspots of the $\Rbasis{3}$ shape
	are slightly pinched by the presence of other harmonics at a lower level.
	In table~\ref{table:ortho-approx}, the right-hand columns
	give an approximation to each exact shape, formed from the
	first ten $\Rbasis{n}$.
	We quote the corresponding cosines in
	table~\ref{table:cosines-approx}.
	The approximations are extremely good, resulting in cosines
	in excess of 0.99.

	The significance of this analysis is not that the
	$S_H$-shape can be roughly
	matched to an element of some complete, orthogonal basis of shapes.
	Such a basis always exists.
	Given a set of trial shapes,
	which could presumably be generated by considering arbitrarily
	exotic higher-derivative operators in the Lagrangian,
	this basis
	could be constructed precisely by
	the Gram--Schmidt procedure described in~\S\ref{sec:new_shape}.
	It is more interesting that,
	at least for the low-dimension operators we are considering,
	the $\Rbasis{n}$ basis
	provides an approximate match
	to the outcome of this process.
	Were we to continue adding new high-dimensional operators
	to the Lagrangian, the $\Rbasis{n}$ shapes presumably
	give guidance about the shapes which could be expected to
	emerge from the Gram--Schmidt procedure.

	\subsection{Distinguishing models}
	These properties imply that, instead of obtaining orthogonal
	combinations from the terms in the Lagrangian as
	in~\S\ref{sec:new_shape},
	it may be possible to do just as well with the $\Rbasis{n}$
	themselves.
	
	Taken at face value, the common appearance of the shape
	in table~\ref{table:ortho-approx} suggests
	that the \emph{shape} of the bispectrum will not serve as a sensitive
	discriminant of microphysics.
	A significant {\Local} mode will favour dominantly
	local interactions,
	driven by gravitational evolution or the scalar potential,
	whereas a significant {\Equilateral} mode will favour
	strong, higher-derivative self-interactions.
	However, it seems difficult to be more precise.
	Instead of focusing on shapes, it may be
	more profitable to study relations between their
	\emph{amplitudes}
	in order to distinguish among competing scenarios.
	
	\paragraph{Partial-wave amplitudes.}
	To proceed, we define a set of amplitudes
	$\beta_n$ for an arbitrary bispectrum $B$,
	\begin{equation}
		\iprod{B_{k_\ast}}{\Rbasis{n}} \equiv  \beta_n \Ps^2(k_\ast),
		\label{eq:beta-hierarchy}
	\end{equation}
	where $\Ps$ is the dimensionless power spectrum
	of the curvature perturbation.
	We will discuss the scale $k_\ast$ below.
	The $\beta_n$ are similar to the
	amplitudes $\fNL^{\mathrm{eq}}$,
	$\fNL^{\mathrm{orth}}$,
	etc., which are used to place constraints on
	the nongaussian fraction observed in real data.
	Any predictive Lagrangian will depend on only a finite number
	of unknown parameters. If enough $\beta_n$ can be estimated
	from data,
	then Eq.~\eqref{eq:beta-hierarchy} allows these parameters
	to be expressed in terms of measurable quantities.
	The remaining relations in Eq.~\eqref{eq:beta-hierarchy},
	when expressed in terms of these measurable quantities,
	constitute predictions of the theory.
	This is rather analogous to an on-shell renormalization scheme
	in scattering calculations by which one expresses ``observables in terms
	of observables.''
	In inflation one usually speaks of ``consistency equations''
	\cite{Copeland:1993jj,Copeland:1993zn}.

	In practice the precise $\beta_n$ depend on the definition of
	the inner product, and indeed will vary between experiments.
	To perform a satisfactory analysis, one should obtain survey-dependent
	predictions for the $\beta_n$.
	The primordial bispectrum should be propagated to the surface of
	last scattering and projected on to the sky,
	and the $\beta_n$ should be computed in the resulting
	two-dimensional inner product.
	The set of basis shapes
	should be orthogonal when measured using the experiment in question,
	and may not be directly related to the $\Rbasis{n}$.
	This will lead to numerically different $\beta_n$ for each survey.
	
	In what follows, we work
	for illustrative purposes with
	the primordial, three-dimensional
	bispectrum rather than the projected bispectrum.
	We make a number of simplifications.
	We use the inner product~\eqref{eq:inner-prod-def}
	in a scale-invariant approximation.%
		\footnote{Our definition coincides with
		Ref.~\cite{Burrage:2011hd}, in which a detailed discussion is given.}
	In evaluating $\iprod{B}{\Rbasis{n}}$
	one must choose a reference or `pivot' scale at which to define
	$B$.
	We have denoted this scale $k_\ast$.
	The bispectrum then contains scale-dependent logarithms of the
	form $\ln k/k_\ast$,
	making $\iprod{B}{\Rbasis{n}}$ a function of $k_\ast$.
	The power spectrum on the right-hand side of Eq.~\eqref{eq:beta-hierarchy}
	is to be evaluated at $k_\ast$.
	Because our implementation of the inner product
	does not retain scale information, we cannot apply this prescription
	precisely. We replace such logarithms
	by $\ln k/k_t$, where $k_t = k_1 + k_2 + k_3$
	is the total scalar 3-momentum.
	This is likely to make an impact on our numerical coefficients
	at next-order, which should therefore be considered approximate.
	
	\paragraph{Example: DBI inflation.}
	As an illustration, we consider DBI inflation governed by
	the action
	\begin{equation}
		S = \int \d^4 x \; \sqrt{-g} \left(
			- \frac{1}{f(\phi)} \left[
				\sqrt{1 - f(\phi) X} - 1
			\right] - V(\phi)
		\right) ,
		\label{eq:dbi}
	\end{equation}
	where $X = -g^{ab} \partial_a \phi \partial_b \phi$.
	This is a simple action in the Horndeski class.
	Based on a microscopic interpretation of~\eqref{eq:dbi}
	as the action for a brane of constant tension
	falling in a warped throat,
	the function $f(\phi)$ is known as the warp factor.
	The potential is $V(\phi)$, and we define
	$\gamma \equiv (1 - f \dot{\phi}^2)^{-1/2}$.
	This action is known to lead to strong nongaussianities
	if $\gamma \gtrsim 1$ \cite{Alishahiha:2004eh,Chen:2006nt}.
	The inflationary fluctuations depend on the
	parameters~\cite{Franche:2009gk}
	\begin{equation}
		\epsilon = \frac{1}{2} \left( \frac{V'}{V} \right)^2 ,
		\quad
		\eta = \frac{V''}{V} ,
		\quad
		\text{and}
		\quad
		\Delta = \sgn(\dot{\phi} f^{1/2}) \frac{f'}{f^{3/2}} \frac{1}{3H} \ ,
		\label{eq:dbi-sr}
	\end{equation}
	where primed quantities are differentiated with respect to $\phi$.
	These must typically be small.
	The bispectrum was determined to $\Or(\epsilon, \eta, \Delta)$
	in Ref.~\cite{Burrage:2011hd}.%
		\footnote{To this accuracy one must typically retain
		gravitational interactions, which are often subdominant
		in models where the bispectrum has significant amplitude.
		Working in the uniform curvature slicing,
		a typical component of the metric is the perturbed lapse,
		$\delta g_{00} \sim \varepsilon \zeta$.
		At quadratic order this will enter
		via an operator such as $(\partial \phi)^2$.
		The leading quadratic operator
		without mixing is $\sim M^2 \dot{\zeta}^2$,
		where $M$ is the mass scale in~\eqref{eq:zeta-action}.
		The leading mixing will be
		roughly $\sim M^2 \varepsilon H \zeta
		\dot{\zeta}$.
		An overdot represents a time derivative, but for this power-counting
		exercise it could be replaced by a generic derivative.
		We estimate the contribution of each operator to a correlation
		function evaluated at characteristic energy scale $E$ by
		setting $\dot{\zeta} \sim E \zeta$. This Minkowski estimate should be
		valid up to horizon exit, where we wish to estimate the relative
		importance of each operator to the density fluctuations which freeze
		in at that time. It follows that
		mixing with the metric can be neglected if
		$E \gtrsim \varepsilon H$.
		
		In Ref.~\cite{Burrage:2010cu}, subleading corrections were
		determined for the covariant Galileon action.
		However, this reference worked in the decoupling limit
		in which mixing with the metric was ignored.
		Typically this will not be consistent, so the
		\emph{quantitative} magnitude of the next-order corrections
		in Ref.~\cite{Burrage:2010cu} should be treated only as a guide.
		In Ref.~\cite{Burrage:2011hd},
		whose results we rely on above, the mixing with the metric
		was retained.}
	
	The Lagrangian depends on the parameters
	of Eq.~\eqref{eq:dbi-sr} and $\gamma$.
	We will therefore require \emph{four} observables
	to fix these parameters. A \emph{fifth} observable enables the theory
	to be tested. The presently well-measured parameters
	are only the amplitude, $\Ps$, and tilt, $n_s$,
	of the scalar power spectrum.
	There are relatively weak constraints on a few modes of the bispectrum.
	In the future it may be possible to detect the tensor amplitude
	$\Ps_g$.
	Assuming it will eventually be possible to measure
	$\beta_0$ and $\beta_1$ together with the tensor-to-scalar
	ratio, $r \equiv \Ps_g / \Ps$,
	then
	using the results of Ref.~\cite{Burrage:2011hd}
	and assuming at least moderate $\gamma$ we find
	\begin{equation}
		\left(2.88 \frac{\beta_1}{\beta_0} - 1\right)
			=
			1.93 (n_s - 1)
			+ 0.03 r \sqrt{-\beta_0}
			+ 2.87 \left(
				6.60 \frac{\beta_2}{\beta_0} + 1
			\right) .
		\label{eq:dbi-consistency-r}
	\end{equation}
	Note that the DBI model predicts $\beta_0 < 0$
	if the bispectrum is large enough to be observable,
	as we will explain below.
	If $r$ cannot be measured, or only with poor accuracy, then it
	will be necessary to use $\beta_3$ as a substitute. In this case,
	we find
	\begin{equation}
		\left(2.88 \frac{\beta_1}{\beta_0} - 1\right)
			=
			0.65 (n_s - 1)
			- 0.02 \left(
				6.60 \frac{\beta_2}{\beta_0} + 1
			\right)
			- 0.17 \left(
				34.98 \frac{\beta_3}{\beta_0} + 1
			\right) .
		\label{eq:dbi-consistency-beta3}
	\end{equation}
	In writing
	Eqs.~\eqref{eq:dbi-consistency-r}--\eqref{eq:dbi-consistency-beta3}
	we must recall that observables (such as the $\beta_n$)
	may mix Lagrangian \emph{parameters}
	at lowest-order, next-order or other higher orders.
	The $\beta_n$ begin at lowest-order, whereas $n_s - 1$
	and $r$ begin at next-order.
	Therefore, in constructing
	\eqref{eq:dbi-consistency-r}--\eqref{eq:dbi-consistency-beta3}
	we have assumed
	\begin{equation}
		\left|
			2.88 \frac{\beta_1}{\beta_0} - 1
		\right|
		\sim
		\left|
			6.60 \frac{\beta_2}{\beta_0} + 1
		\right|
		\lesssim
		|n_s - 1|
		\sim
		r .
		\label{eq:hierarchy}
	\end{equation}
	
	Whether Eq.~\eqref{eq:dbi-consistency-r}
	or~\eqref{eq:dbi-consistency-beta3}
	is more useful depends on the relative difficulty of measuring
	$r$ and $\beta_3$.
	These expressions constitute a \emph{model-independent} test of the
	DBI framework: they hold for any
	DBI action, up to
	$\Or(\epsilon, \eta, \Delta)$, no matter what potential or warp factor
	is chosen.
	By
	showing they are not satisfied,
	one could rule out the DBI action as the origin of the inflationary
	perturbations.
	Of course, there is not a one-to-one mapping from models
	to consistency relations such
	as~\eqref{eq:dbi-consistency-r}--\eqref{eq:dbi-consistency-beta3},
	and determining that any such equation \emph{is} satisfied
	does not provide decisive evidence in favour of a model.
	The utility of such equations lies with their power to rule models
	out.
	However, 
	showing that the $\beta_n$ satisfy
	a hierarchy of consistency equations derived from some
	Lagrangian would be circumstantial evidence in favour of
	that model, especially if the agreement could be shown
	to persist to large $n$.
	
	Eqs.~\eqref{eq:dbi-consistency-r}--\eqref{eq:dbi-consistency-beta3}
	are analogues of the ``next-order'' consistency equations
	for the tensor tilt, $n_t$ (cf. Eqs. (5.6)--(5.7) of
	Lidsey~{\etal}~\cite{Lidsey:1995np}).
	If the $\beta_n$ cannot be determined with sufficient accuracy
	to test these equations, we can obtain a simpler set of ``lowest--order''
	relations obtained by systematically neglecting next-order terms,
	which entails $n_s - 1 \approx r \approx 0$.
	Together with~%
	\eqref{eq:dbi-consistency-r}--\eqref{eq:dbi-consistency-beta3},
	Eq.~\eqref{eq:hierarchy} then implies
	\begin{equation}
		2.88 \frac{\beta_1}{\beta_0} \approx
		-6.60 \frac{\beta_2}{\beta_0} \approx 1 .
		\label{eq:dbi-simple}
	\end{equation}
	Even more simply, Eq.~\eqref{eq:dbi-simple}
	requires $\beta_0$ and $\beta_1$ to have the same sign,
	and $\beta_2$ to have the opposite sign.
	By consulting the individual expressions for the $\beta_n$,
	it follows that $\beta_0$ and $\beta_1$ must be negative
	but $\beta_2$ should be positive
	whenever $\gamma$ is moderately large.
	This test is applicable even if the $\beta_n$ cannot be
	determined accurately.
	In the present framework,
	it is a manifestation of the well-known
	result that the DBI model produces $\fNL^{\mathrm{eq}} < 0$,
	whereas WMAP data favour $\fNL^{\mathrm{eq}} \gtrsim 0$.
	For this reason, present-day observations are sufficient to disfavour
	DBI inflation.
	Note that
	Eq.~\eqref{eq:dbi-simple},
	and similar expressions for
	$\beta_n$ with $n > 2$, express the expected
	decrease in amplitude of $\iprod{B}{\Rbasis{n}}$
	with increasing $n$.
	The decrease is not monotonic, because the spikes which appear in
	$\Rbasis{n}$ at larger $n$ cause a small enhancement.
	However, the $n = 0,1,2$ harmonics are larger than the rest,
	which is required by the analysis of \S\ref{sec:new_shape}.

	\paragraph{Example: $k$-inflation.}
	For comparison, consider the power-law $k$-inflation model of
	Armend\'{a}riz-Pic\'{o}n {\etal}~\cite{ArmendarizPicon:1999rj}.
	The action for this model satisfies
	\begin{equation}
		S = \int \d^4 x \, \sqrt{-g} \; \frac{4}{9} \frac{4-3\gamma}{\gamma^2}
			\frac{X^2 - X}{\phi^2} .
	\end{equation}
	It admits an inflationary solution for $X=(2-\gamma)/(4-3\gamma)$ 
	provided $0 < \gamma < 2/3$.
	(Note that $\gamma$ in this model is just a parameter,
	not related to the $\gamma$ of the DBI model.)
	In the
	limit $\gamma \ll 1$, and
	keeping only leading-order terms, this model predicts
	\begin{equation}
		2.61 \frac{\beta_1}{\beta_0} =
		- 4.80 \frac{\beta_2}{\beta_0} = 1 .
		\label{eq:k-simple}
	\end{equation}
	Comparison with~\eqref{eq:dbi-simple} shows that it
	would be necessary to measure $\beta_0 / \beta_1$ to
	about $10\%$ in order to distinguish these models.
	A sufficiently accurate measurement of $\beta_2$
	would make the test considerably easier to apply.
	
	This method is closely related to a
	trispectrum-based
	test for single-field inflation
	proposed by Smidt {\etal}~\cite{Smidt:2010ra}.
	The trispectrum contains contributions from two different
	`{\Local}' shapes, with amplitudes parametrized by $\tauNL$
	and $\gNL$
	\cite{Sasaki:2006kq,Byrnes:2006vq,Seery:2006js}.
	The $\tauNL$ contribution obeys the
	Suyama--Yamaguchi inequality $\tauNL \geq (6 \fNL^{\mathrm{local}} / 5)^2$
	\cite{Suyama:2007bg,Smith:2011if}.
	Smidt~{\etal} suggested studying
	$A = \tauNL / (6 \fNL^{\mathrm{local}}/5)^2$,
	which is analogous to the ratios $\beta_n / \beta_0$
	introduced above.
	Their analysis suggested that Planck may be able to
	measure $A$ to $\pm 1.0$ at $1\sigma$, and a future CMB satellite
	may even be able to achieve $\pm 0.3$
	with the same significance.
	An accurate measurement of $A > 1$ would be sufficient to rule out
	single-field scenarios.

	Like the well-known
	standard inflationary consistency relation,
	whether relationships
	such as~\eqref{eq:dbi-consistency-r}--\eqref{eq:dbi-simple}
	and~\eqref{eq:k-simple}
	are useful in practice will depend
	on the accuracy with which each component can be measured.
	This depends on the signal-to-noise associated with each
	shape.
	However, the method we have described can be implemented
	with \emph{any} suitable basis; it is not restricted to the
	$\Rbasis{n}$ functions described above.

\section{Conclusions}
\label{sec:conclusions}

	Whichever microphysics operated
	in the very early Universe, its remnants are
	encoded in the CMB radiation.
	The imminent arrival of \emph{Planck}
	data
	will enable us to assemble a detailed picture of
	the microwave sky,
	accompanied by important information
	concerning the statistics of the temperature
	and polarization fields.
	Searching for non-gaussianities in these statistics is
	a promising strategy to determine the details of
	interactions during the inflationary era.

	In this paper, we have
	revisited the bispectrum in very general models of
	single-field inflation,
	which has recently been obtained by
	Gao \& Steer \cite{Gao:2011qe} 
	(see also Renaux-Petel~\cite{RenauxPetel:2011sb})
	and de Felice \& Tsujikawa \cite{DeFelice:2011uc}.
	These computations demonstrated
	that, even in very general scenarios, the inflationary fluctuations
	would be generated by the same Lagrangian operators which are
	present in much simpler scenarios such as $k$-inflation.
	The difference lies only in the pattern of
	correlations among their coefficients, which varies between scenarios.
	We have shown that, although a potentially distinctive shape
	can be generated by these generalized models,
	it requires a degree of fine-tuning
	(or a large overall bispectrum).
	In any case, this shape is similar to one which has been
	encountered elsewhere
	\cite{Creminelli:2010qf,Burrage:2011hd}.
	We conclude that it will be very difficult to distinguish between
	single-field models purely by detecting shapes in the bispectrum,
	although useful qualitative guidance could perhaps be obtained.
	
	The natural alternative is to study correlations among the
	amplitudes of shapes which \emph{are} present.
	For this purpose one may use templates or decompose
	the bispectrum into an orthogonal basis.
	For illustration, we use a basis proposed by Fergusson~{\etal}
	\cite{Fergusson:2008ra,Fergusson:2009nv}.
	A given Lagrangian will typically generate fluctuations which
	depend on a finite number of parameters. If enough
	modes of the bispectrum can be determined with sufficient accuracy,
	these parameters can be written in terms of observable
	quantities. Further observations then constitute
	tests of any particular model.
	
	As an illustration, we have applied our method
	to DBI inflation with an arbitrary potential and warp factor,
	and compared with the $k$-inflation scenario.
	With sufficiently accurate observations it may be possible
	to distinguish these scenarios.
	However, 
	similar tests can be devised for any single-field
	inflationary model.

	\acknowledgments
	We would like to thank Daniel Baumann, Clare Burrage, Anne Davis,
	James Fergusson, Eugene Lim,
	Donough Regan and S\'{e}bastien Renaux-Petel
	for helpful discussions.
	RHR is supported by Funda\c{c}\~{a}o para a Ci\^{e}ncia e a Tecnologia 
	through the grant SFRH/BD/35984/2007 and acknowledges 
	the hospitality of the University of Sussex
	whilst this work was being completed.
	DS was supported by the Science and Technology Facilities Council
	[grant number ST/F002858/1].

	\appendix

	\section{Shape functions---different parametrizations}

	In plotting the bispectrum shapes we have used two parametrizations,
	which we describe in what follows.
	\paragraph{Babich \etal}
	This consists in factorizing 
	one of the wavenumbers, say $k_3$, in the bispectrum amplitude, 
	and rescale the independent remaining momenta accordingly, such that
	$0 \leq k_1/k_3, k_2/k_3 \leq 1$. 
	The shape function is given by
	\begin{equation*}
		\Big(\dfrac{k_1}{k_3} \Big)^2
		\Big(\dfrac{k_2}{k_3} \Big)^2
		B\bigg(\dfrac{k_1}{k_3}, \dfrac{k_2}{k_3}, 1  \bigg) .
	\end{equation*}

	\paragraph{Fergusson \& Shellard.}
	In this parametrization
	the privileged momentum scale is given by the semi-perimeter of the
	triangular configuration, used
	to define new variables $\alpha$ and $\beta$, which satisfy
	\begin{equation*}
		k_1 = \dfrac{k_t}{4} (1+\alpha+\beta) , \quad
		k_2 = \dfrac{k_t}{4} (1-\alpha+\beta) , \quad \text{and} \quad 
		k_3 =	\dfrac{k_t}{2} (1-\beta) .
	\end{equation*}
	The range of domain of $\beta$ is $0\leq \beta \leq 1$, 
	whereas $\beta-1 \leq \alpha \leq 1-\beta$.
	The shape function is given by the combination
	\begin{equation*}
		k_1^2 k_2^2 k_3^2 B\big(k_1 , k_2 , k_3\big) .
	\end{equation*}	

\begin{sidewaystable}
	
	\tablepreamble

   \sbox{\boxplot}{%
   	\includegraphics[scale=0.15]{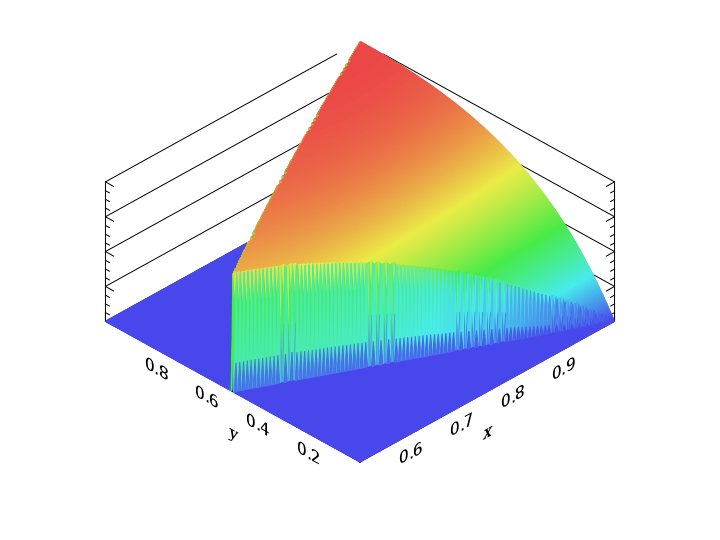}
   }
   \settowidth{\plotw}{\usebox{\boxplot}}
   \sbox{\boxplota}{%
   	\includegraphics[scale=0.1]{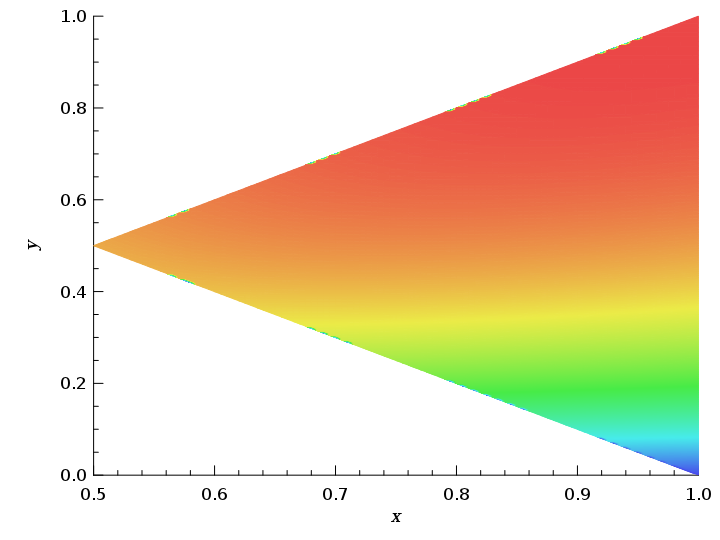}
   }
   \settowidth{\plotwa}{\usebox{\boxplota}}

   \sbox{\tableA}{%
		\begin{tabular}{ccccc}

			\toprule

			&
			\multicolumn{2}{c}{Babich {\etal}} &
			\multicolumn{2}{c}{Fergusson \& Shellard}
			\\
			
			\cmidrule(r){2-3}
			\cmidrule(l){4-5}
			
			$S_{\zeta'^3}$ &
			\parbox[c]{\plotwa}{\includegraphics[scale=0.1]{Plots/Babich/2D/ShapeLO/Lambda1}} &
			\parbox[c]{\plotw}{\includegraphics[scale=0.15]{Plots/Babich/3D/ShapeLO/Lambda1}} &
			\parbox[c]{\plotwa}{\includegraphics[scale=0.1]{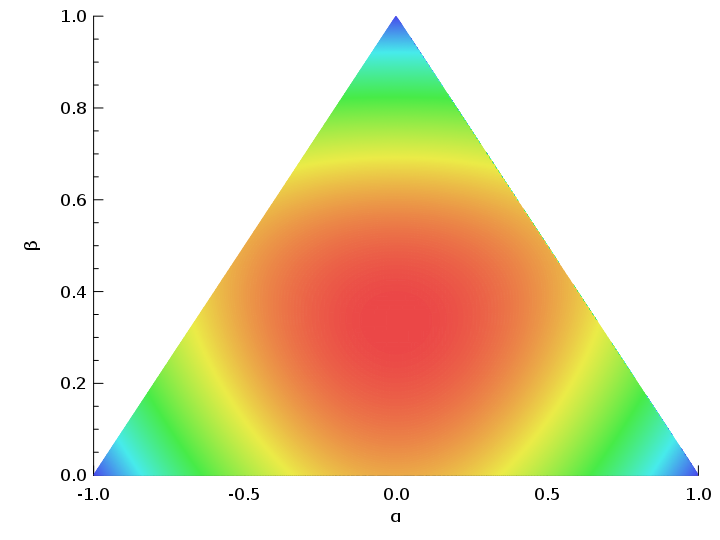}} &
			\parbox[c]{\plotw}{\includegraphics[scale=0.15]{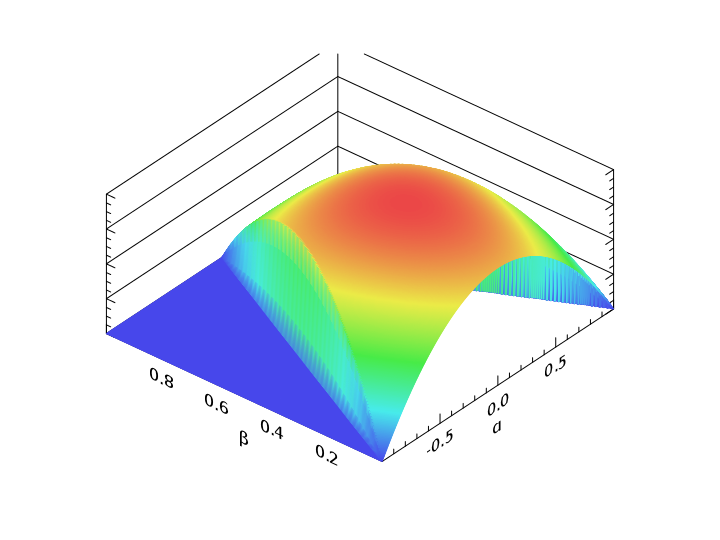}}
			\\

	$S_{\zeta \zeta'^2}$ &
			\parbox[c]{\plotwa}{\includegraphics[scale=0.1]{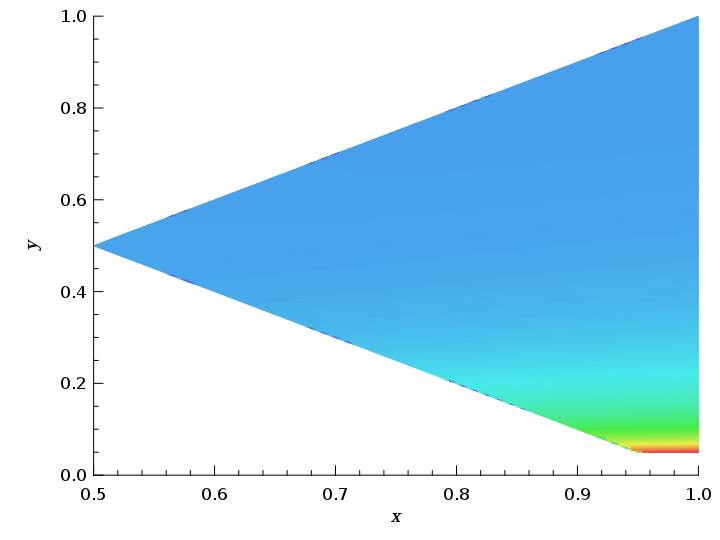}} &
			\parbox[c]{\plotw}{\includegraphics[scale=0.15]{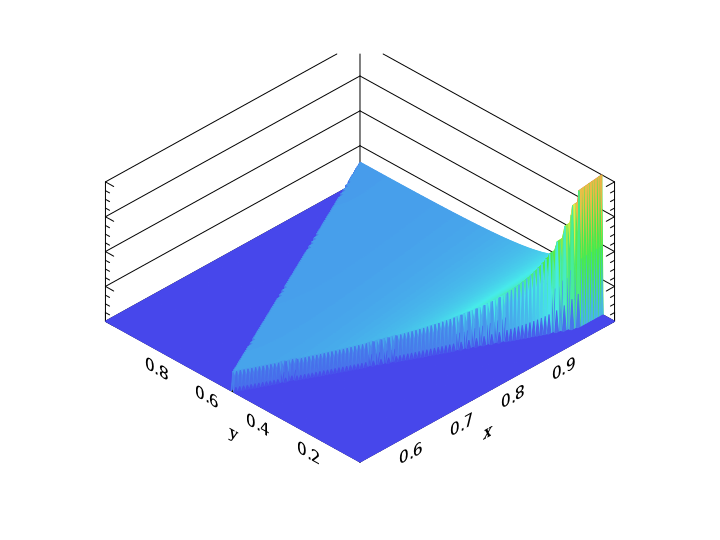}} &
			\parbox[c]{\plotwa}{\includegraphics[scale=0.1]{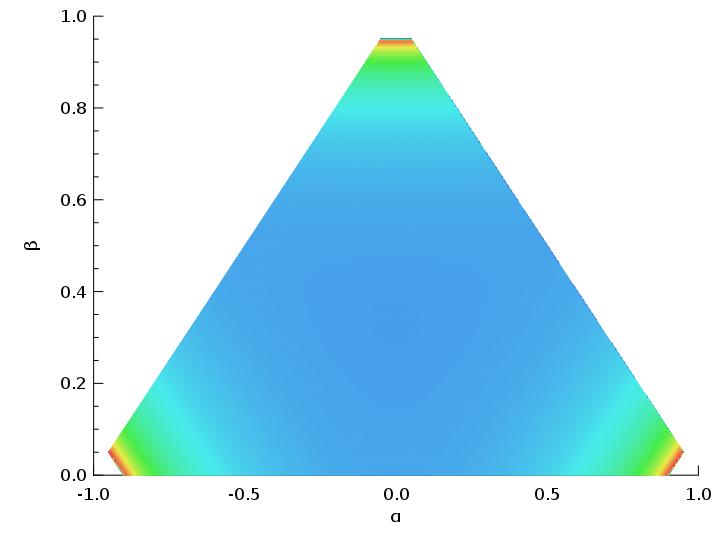}} &
			\parbox[c]{\plotw}{\includegraphics[scale=0.15]{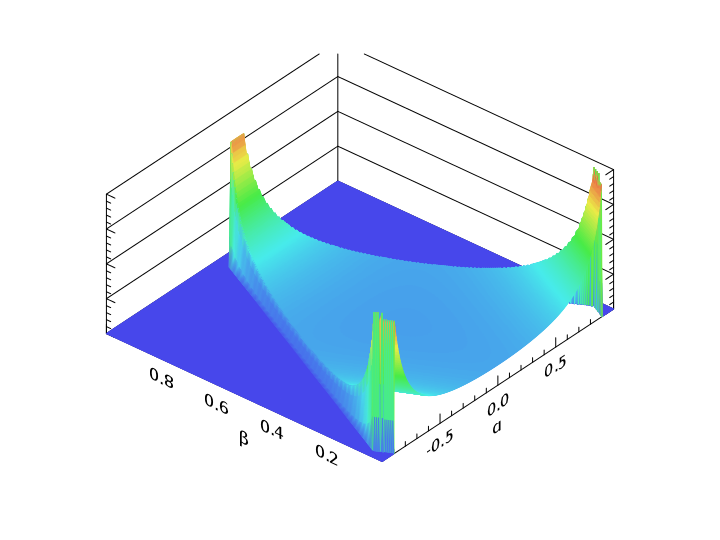}}
			\\
			
			$S_{\zeta (\partial \zeta)^2}$ &
			\parbox[c]{\plotwa}{\includegraphics[scale=0.1]{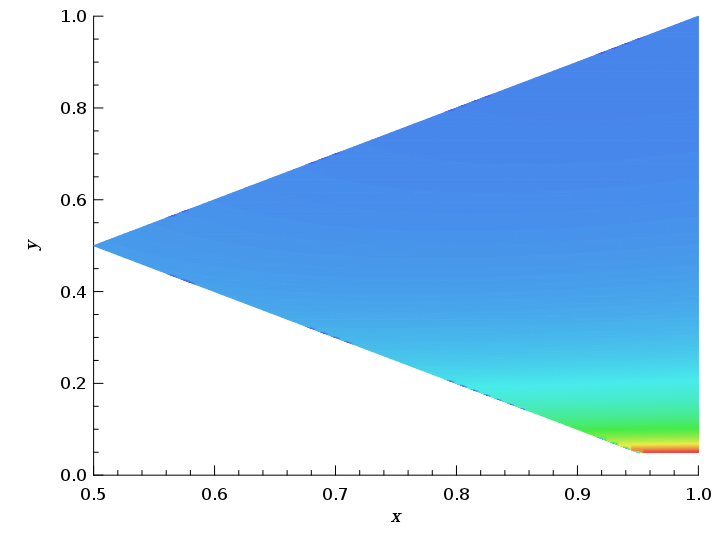}} &
			\parbox[c]{\plotw}{\includegraphics[scale=0.15]{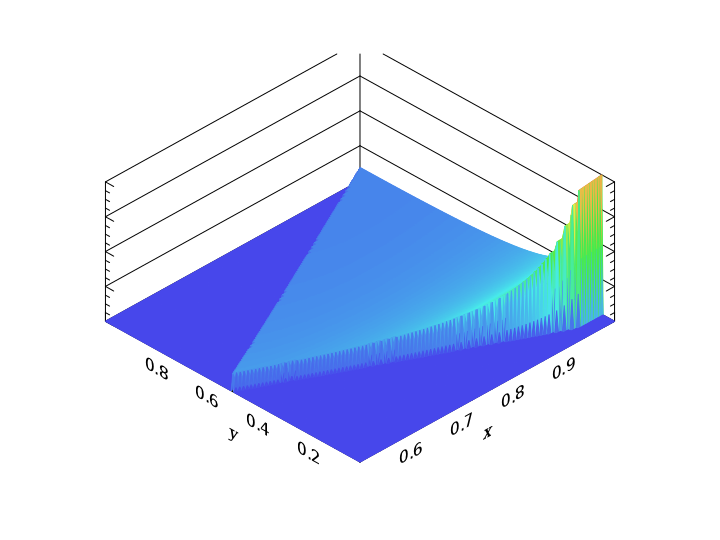}} &
			\parbox[c]{\plotwa}{\includegraphics[scale=0.1]{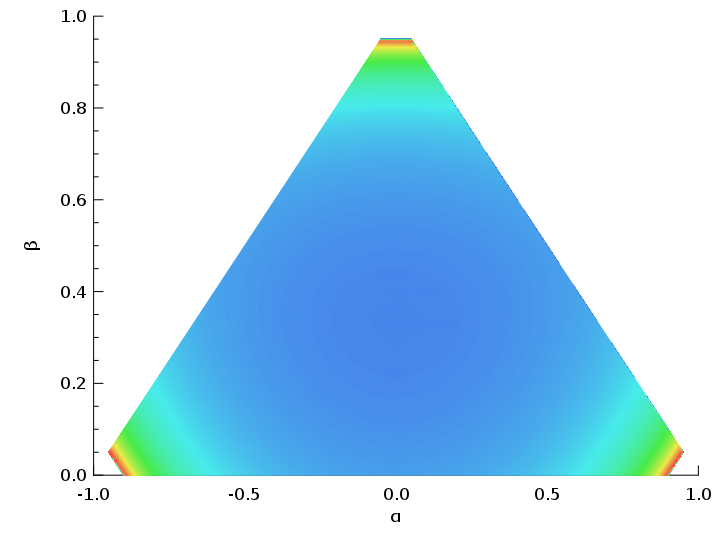}} &
			\parbox[c]{\plotw}{\includegraphics[scale=0.15]{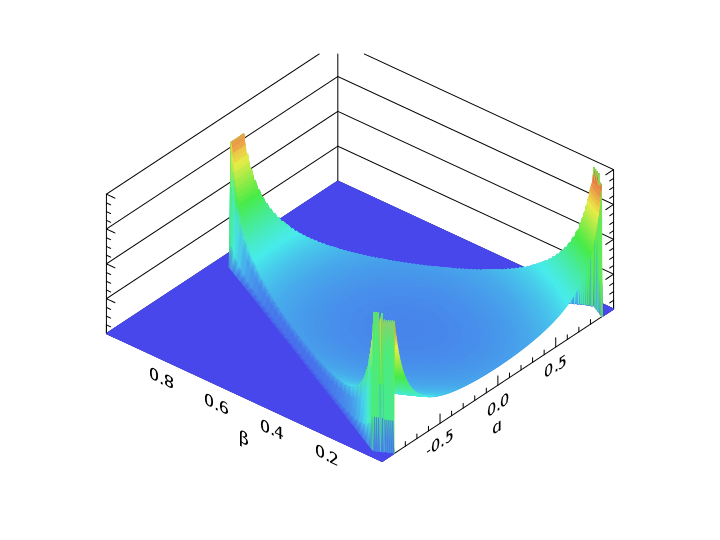}}
			\\

			$S_{ \zeta' \partial_i \zeta \partial^{i} (\partial^{-2} \zeta)}$ &
			\parbox[c]{\plotwa}{\includegraphics[scale=0.1]{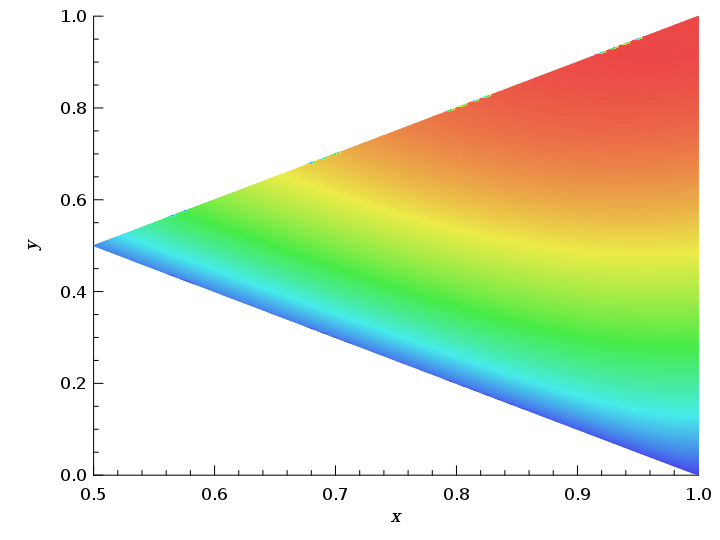}} &
			\parbox[c]{\plotw}{\includegraphics[scale=0.15]{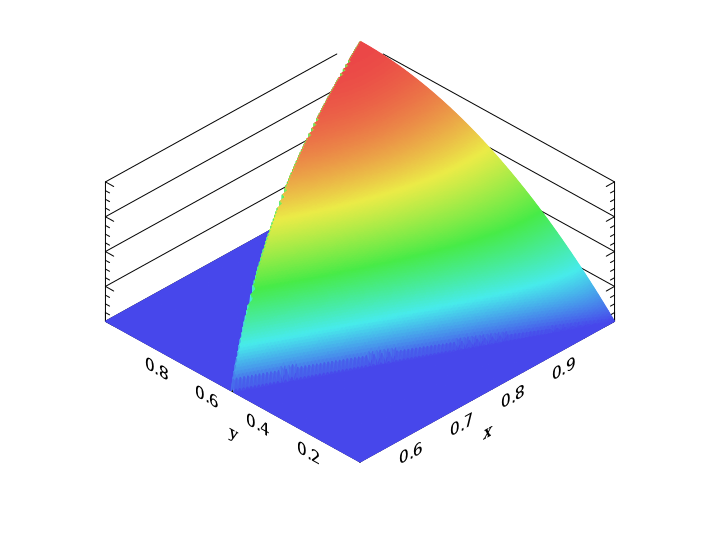}} &
			\parbox[c]{\plotwa}{\includegraphics[scale=0.1]{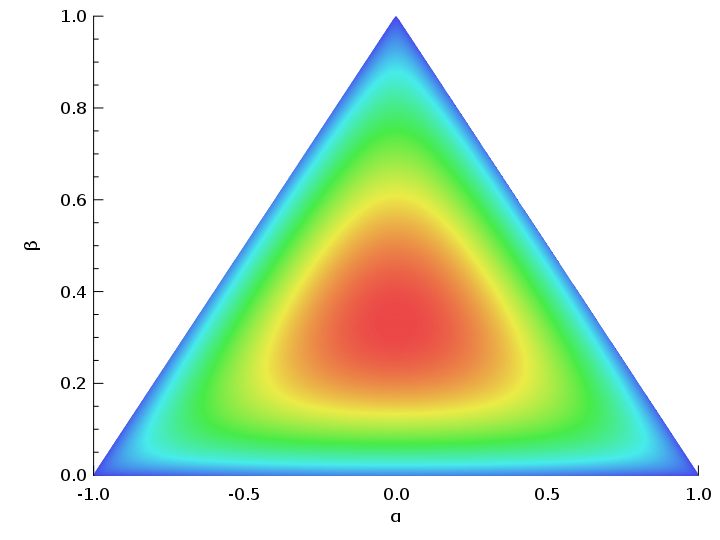}} &
			\parbox[c]{\plotw}{\includegraphics[scale=0.15]{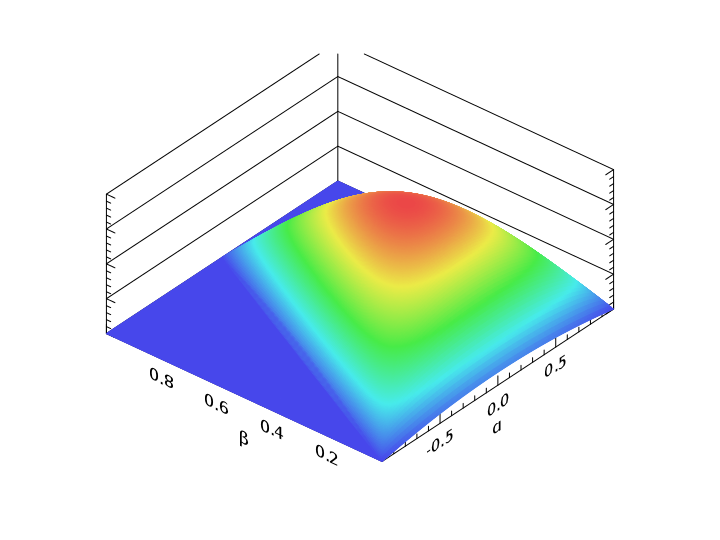}}\\
			
			$S_{ \partial^2 \zeta (\partial_i \partial^{-2}\zeta')^2}$ &
			\parbox[c]{\plotwa}{\includegraphics[scale=0.1]{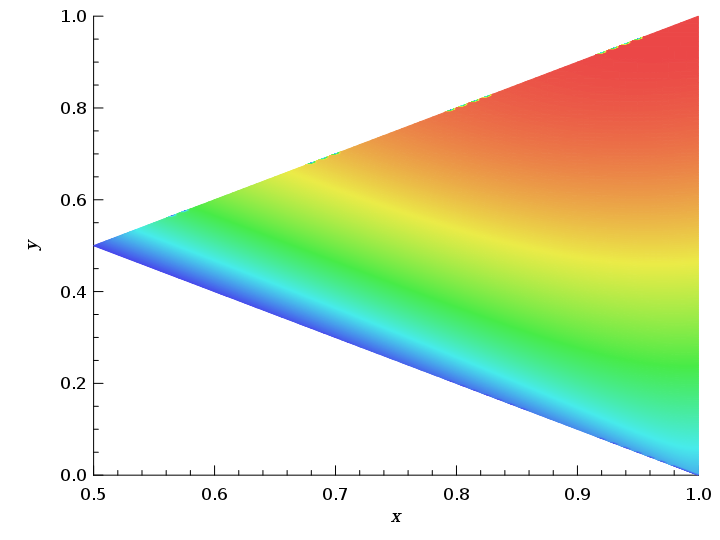}} &
			\parbox[c]{\plotw}{\includegraphics[scale=0.15]{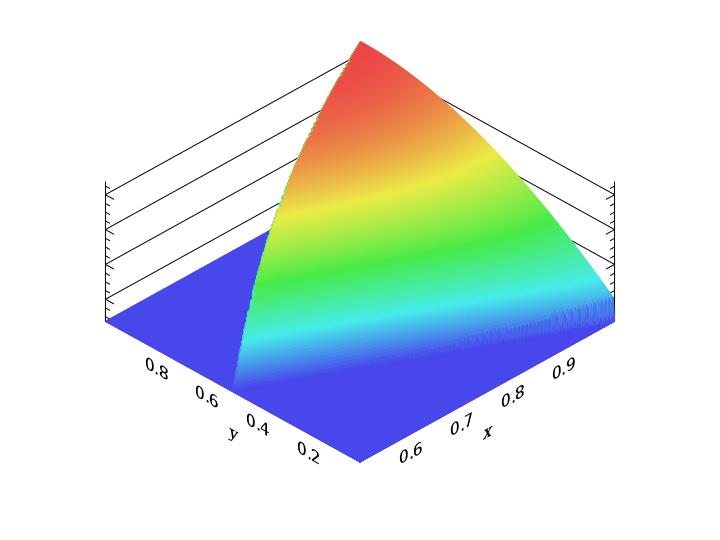}} &
			\parbox[c]{\plotwa}{\includegraphics[scale=0.1]{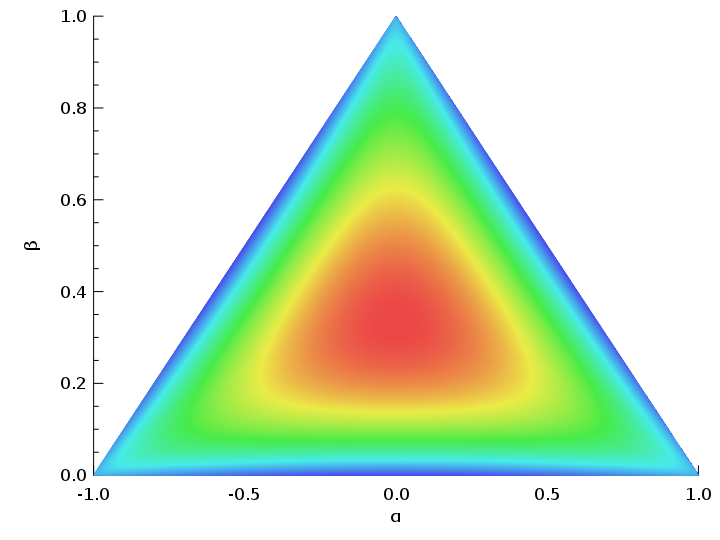}} &
			\parbox[c]{\plotw}{\includegraphics[scale=0.15]{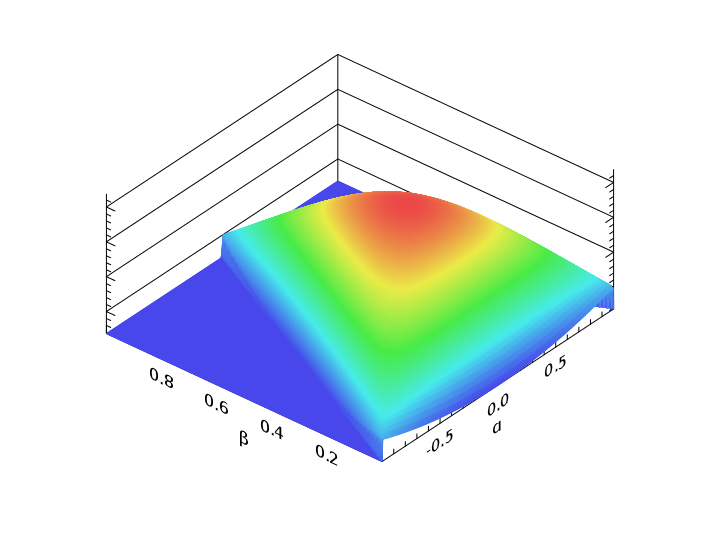}}\\
			\bottomrule

		\end{tabular}
	}
	\settowidth{\tblw}{\usebox{\tableA}}
	\addtolength{\tblw}{-1em}

	\begin{center}
		\usebox{\tableA}
	\end{center}
			
	\caption{\label{table:shapes-leading-order}Bispectrum shapes 
	at leading order for the operators in the
	action~\eqref{eq:zeta-action}, using the
	Babich {\etal} \cite{Babich:2004gb}
	and Fergusson-Shellard \cite{Fergusson:2008ra}
	parametrizations (see appendix).}

	\end{sidewaystable}

\begin{table}

	\tablepreamble
		
	\sbox{\tableA}{%
		\begin{tabular}{QqQqQq}
			 	\toprule

			&
			\multicolumn{5}{c}{shapes at leading order}
			\\

			\cmidrule(l){2-6}

		 	&
		 	\multicolumn{1}{c}{$S_{\zeta'^3}$} &
		 	\multicolumn{1}{c}{$S_{\zeta \zeta'^2}$} &
		 	\multicolumn{1}{c}{$S_{\zeta (\partial \zeta)^2}$} &
		 	\multicolumn{1}{c}{$S_{\zeta' \partial_i \zeta \partial^{i} (\partial^{-2}\zeta')}$} &
		 	\multicolumn{1}{c}{$S_{\partial^2 \zeta (\partial_i \partial^{-2}\zeta')^2}$} 
		 	
		 	\\

			\midrule

			\textrm{{\Local}} &

			0.42 &
			0.99 & 
			1.00 & 0.35 & 0.31
			\\

			\cmidrule{2-6}

			\textrm{{\Equilateral}} &
			0.94 	&
			 0.44 &
			0.38 & 1.00 & 0.99
			\\

			\cmidrule{2-6}

			\textrm{{\Orthogonal}} &
				0.29 &
			0.50 &
			0.49 & 0.02 & 0.12
			\\

			\cmidrule{2-6}

			\textrm{{\Enfolded}} &
			0.75 &
			0.65 &
			0.62 & 0.55 & 0.43
			\\
			
				
			


			\bottomrule

		\end{tabular}
	}
	\settowidth{\tblw}{\usebox{\tableA}}
	\addtolength{\tblw}{-1em}

	\begin{center}
		\usebox{\tableA}
	\end{center}

	\renewcommand{\arraystretch}{1}
	
	

	\caption{\label{table:leading_overlaps}Cosines
	between the leading order shapes and the
	common templates used in CMB analysis.}
	
	\end{table}

\vspace*{0.5cm}

	\begin{table}[h]

	\tablepreamble
	
	\sbox{\tableA}{%
		\begin{tabular}{cTcTcT}

			\toprule

				\multicolumn{1}{c}{{\Local}\tmark{a} } &
				\multicolumn{1}{c}{{\Equilateral}} &
				\multicolumn{1}{c}{{\Orthogonal}} &
				\multicolumn{1}{c}{{\Enfolded}} &
				\multicolumn{1}{c}{Creminelli {\etal}\tmark{b}} &
				\multicolumn{1}{c}{$P(X,\phi)$\tmark{c}}
			\\
			\cmidrule{1-6}

				 0.02 &
				 0.00 &
				 0.00 &
				 0.00 &
				 0.99& 0.86
\\
			\bottomrule
		
		\end{tabular}
	}
	\settowidth{\tblw}{\usebox{\tableA}}
	\addtolength{\tblw}{-1em}
	
	\begin{center}
		\usebox{\tableA}
	\end{center}
	
	\renewcommand{\arraystretch}{1.0}
	\scriptsize
	
	\sbox{\tableB}{%
		\begin{tabular}{l@{\hspace{1mm}}l}	
			\tmark{a} & \parbox[t]{\tblw}{%
				The {\Local} template
				is divergent and requires choosing an appropriate regulator.}
				\\
			\tmark{b} & \parbox[t]{\tblw}{%
				This is the shape studied by 
				Creminelli {\etal} \cite{Creminelli:2010qf}.
				}\\
				\tmark{c} & \parbox[t]{\tblw}{%
				This is the shape $O$ constructed at next-order in
				$P(X,\phi)$ models \cite{Burrage:2011hd}.
				For the purpose of comparison, an
				appropriate normalization of the $P(X, \phi)$
				spectrum has been chosen.
				}
		\end{tabular}
	}
	
	\begin{center}
		\usebox{\tableB}
	\end{center}

	\caption{\label{table:orthogonal-cosines}Cosines 
	between the $S_H$-shape~\eqref{eq:o-def} and
	common templates.}
	\end{table}

	\begin{table}[h]
	
	\small
	\heavyrulewidth=.08em
	\lightrulewidth=.05em
	\cmidrulewidth=.03em
	\belowrulesep=.65ex
	\belowbottomsep=0pt
	\aboverulesep=.4ex
	\abovetopsep=0pt
	\cmidrulesep=\doublerulesep
	\cmidrulekern=.5em
	\defaultaddspace=.5em
	\renewcommand{\arraystretch}{1.6}

    \sbox{\boxplot}{%
    	\includegraphics[scale=0.15]{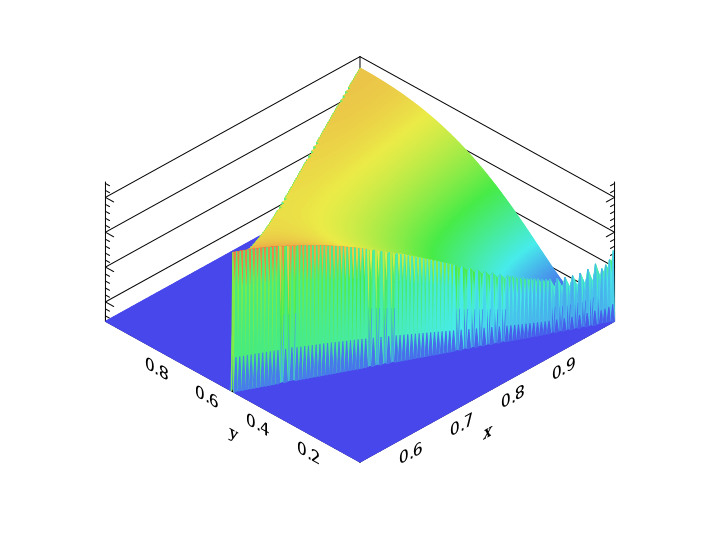}
    }
    \settowidth{\plotw}{\usebox{\boxplot}}
    \sbox{\boxplota}{%
    	\includegraphics[scale=0.1]{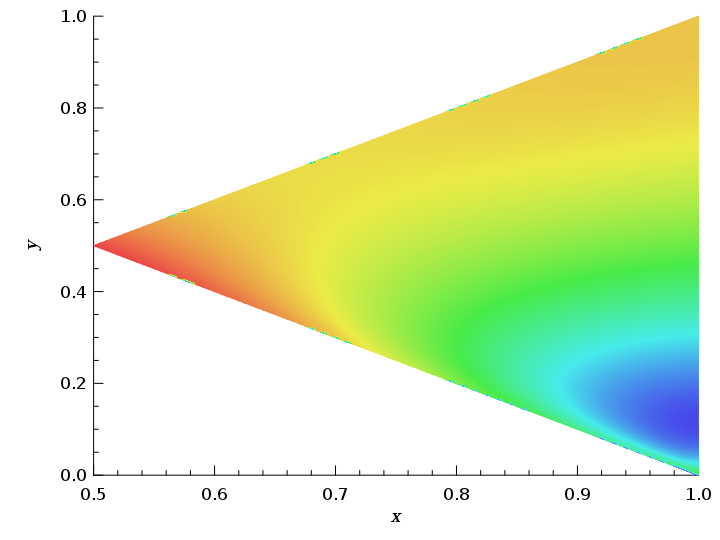}
    }
    \settowidth{\plotw}{\usebox{\boxplota}}
    
    \sbox{\tableA}{%
		\begin{tabular}{ccccc}

			\toprule

			\multicolumn{2}{c}{Shapes } & \multicolumn{1}{c}{} &
			\multicolumn{2}{c}{Approximations} 
			\\
			
			\cmidrule(r){1-2}
			\cmidrule(l){4-5}	
			
			\multicolumn{5}{c}{$S_H$ shape}
			\\			

			
			\parbox[c]{\plotw}{\includegraphics[scale=0.1]{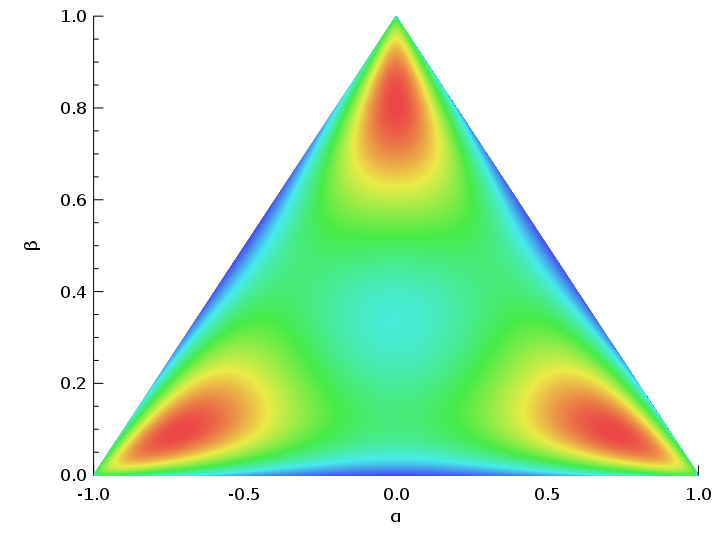}} &
			\parbox[c]{\plotw}{\includegraphics[scale=0.15]{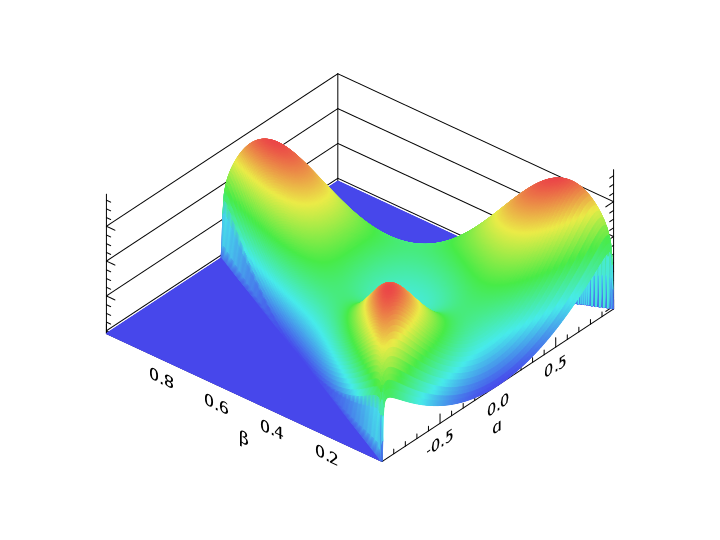}}& &
			\parbox[c]{\plotw}{\includegraphics[scale=0.1]{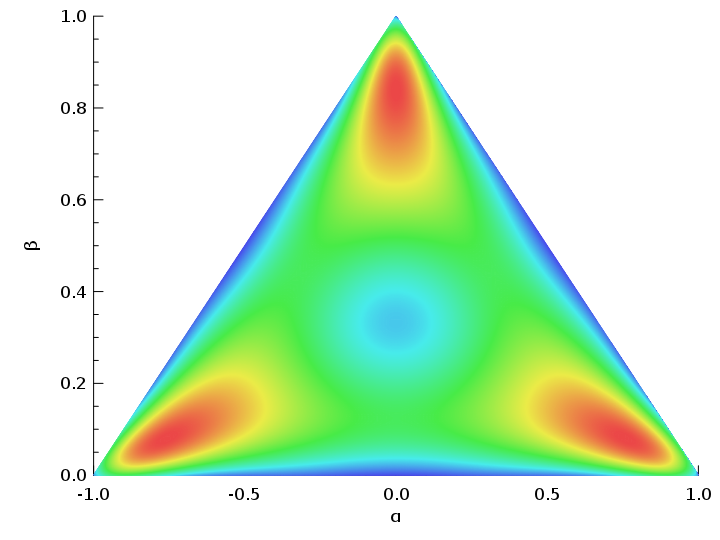}} &
			\parbox[c]{\plotw}{\includegraphics[scale=0.15]{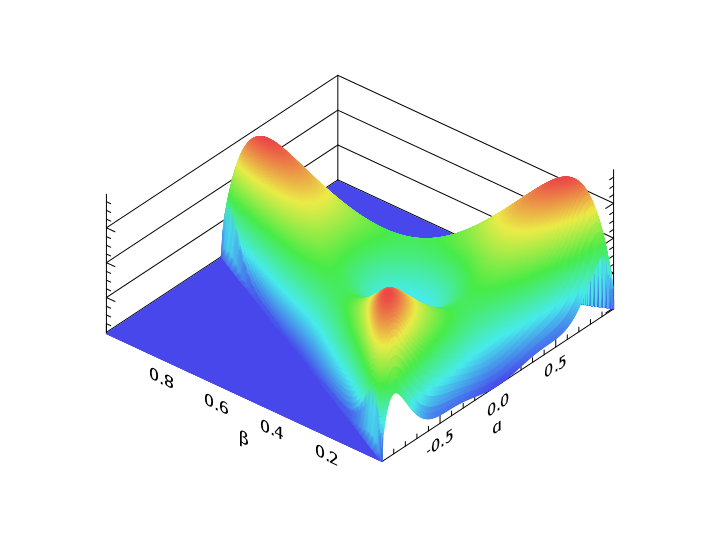}}
			\\

			\midrule
			
			\multicolumn{5}{c}{Creminelli {\etal} orthogonal shape}
			\\
			
			\parbox[c]{\plotw}{\includegraphics[scale=0.1]{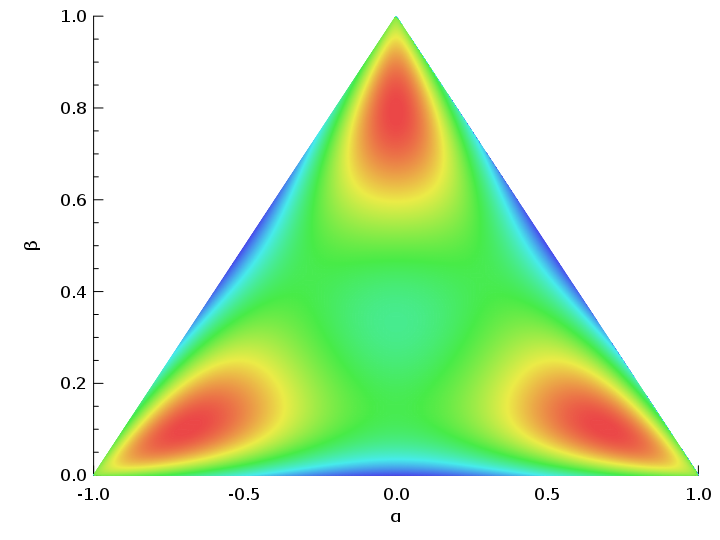}} &
			\parbox[c]{\plotw}{\includegraphics[scale=0.15]{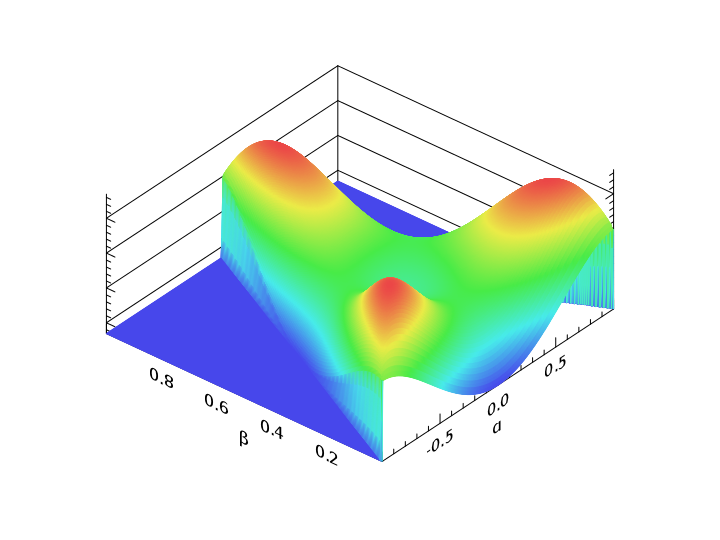}}& &
			\parbox[c]{\plotw}{\includegraphics[scale=0.1]{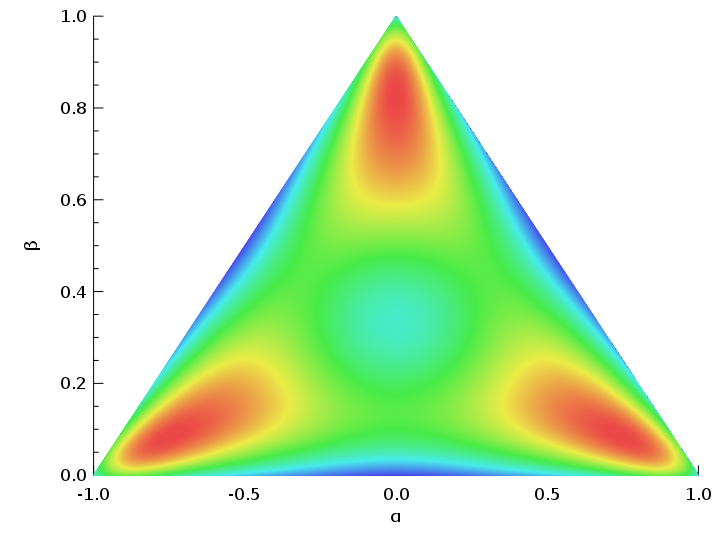}} &
			\parbox[c]{\plotw}{\includegraphics[scale=0.15]{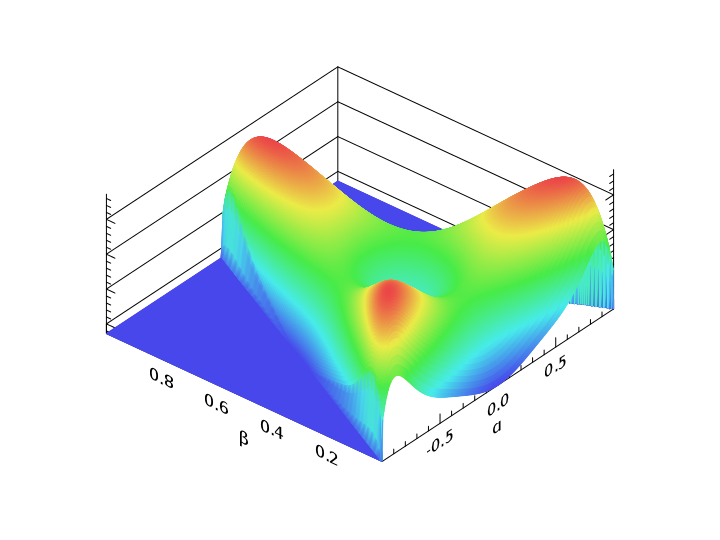}}
			\\
			
			\midrule

\multicolumn{5}{c}{$P(X,\phi)$ orthogonal shape}
			\\
			
				\parbox[c]{\plotw}{\includegraphics[scale=0.1]{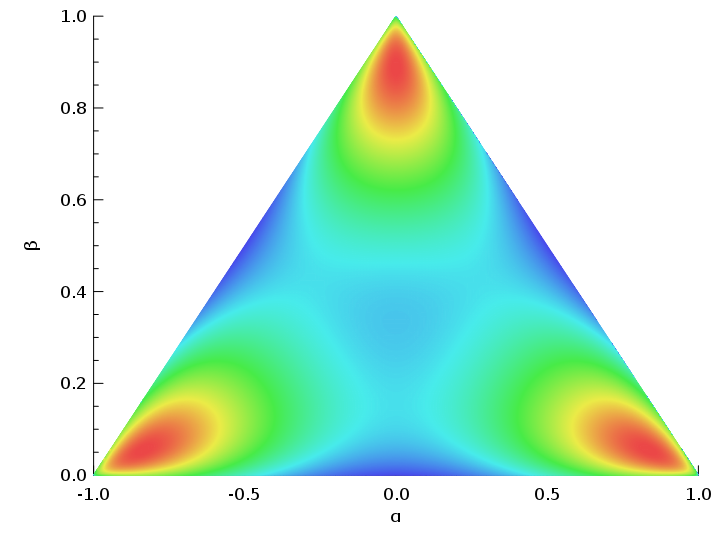}} &
			\parbox[c]{\plotw}{\includegraphics[scale=0.15]{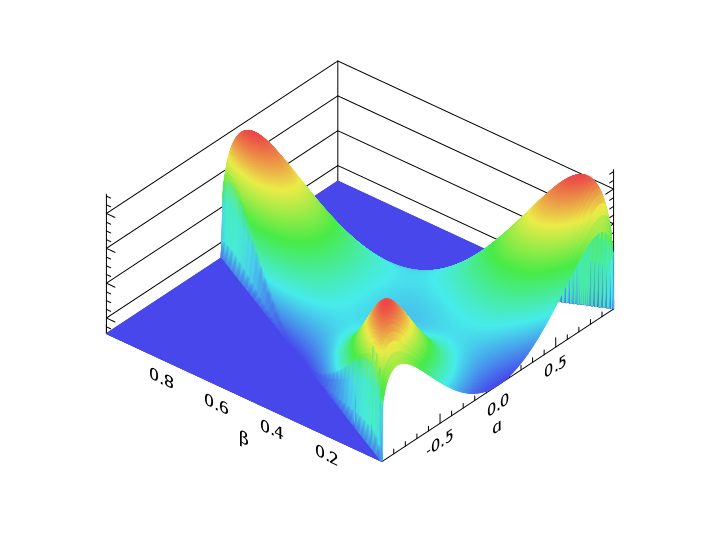}}& &
			\parbox[c]{\plotw}{\includegraphics[scale=0.1]{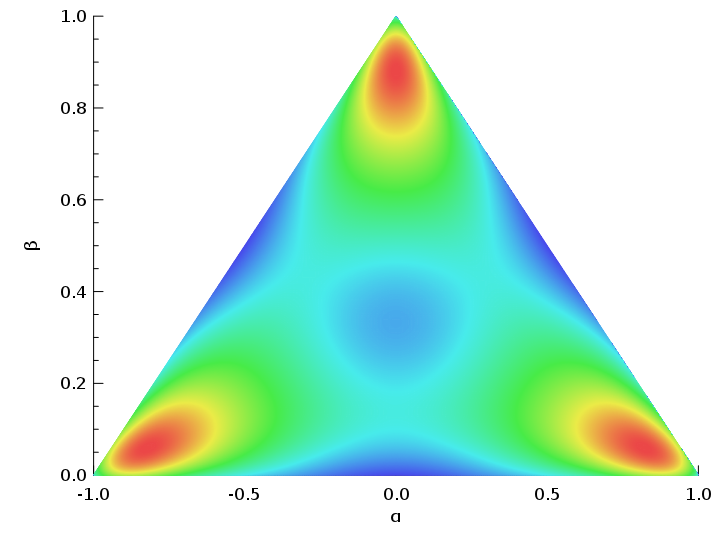}} &
			\parbox[c]{\plotw}{\includegraphics[scale=0.15]{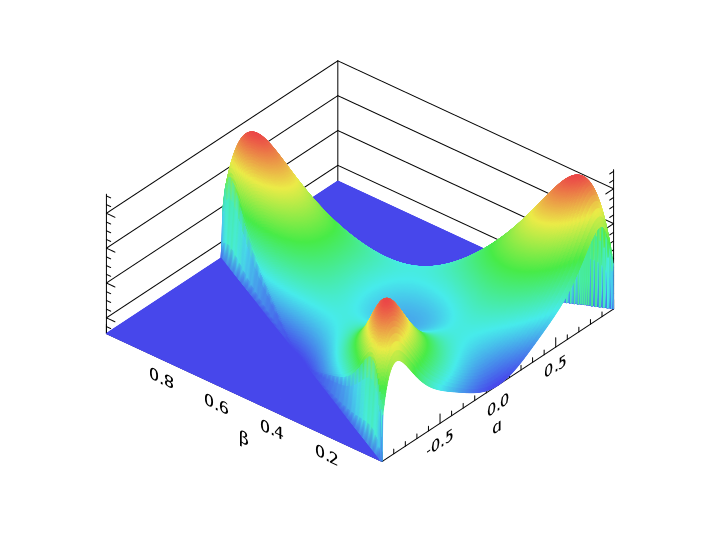}}
			\\

			\bottomrule

		\end{tabular}
	}
	\settowidth{\tblw}{\usebox{\tableA}}
	\addtolength{\tblw}{-1em}

	\begin{center}
		\usebox{\tableA}
	\end{center}
			
	\caption{\label{table:ortho-approx}
	Approximations to the $S_H$-shape and similar bispectra, 
	up to the first ten harmonics in the $\mathcal{R}_n$
	expansion. The coefficients $\alpha_n$ are listed
	in table~\ref{table:Rorthogonal}.
	}

	\end{table}

	\begin{sidewaystable}[h]
	
	\tablepreamble
	
	\sbox{\boxplot}{%
		\includegraphics[scale=0.125]{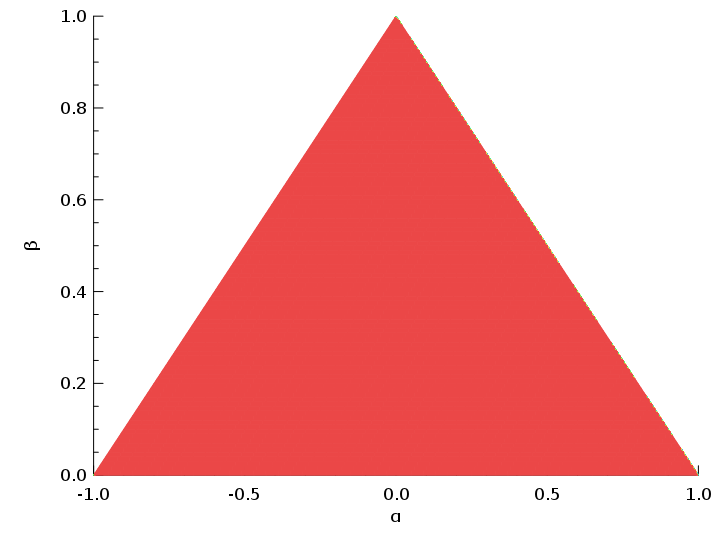}
	}
	\settowidth{\plotw}{\usebox{\boxplot}}
	\sbox{\boxplota}{%
		\includegraphics[scale=0.18]{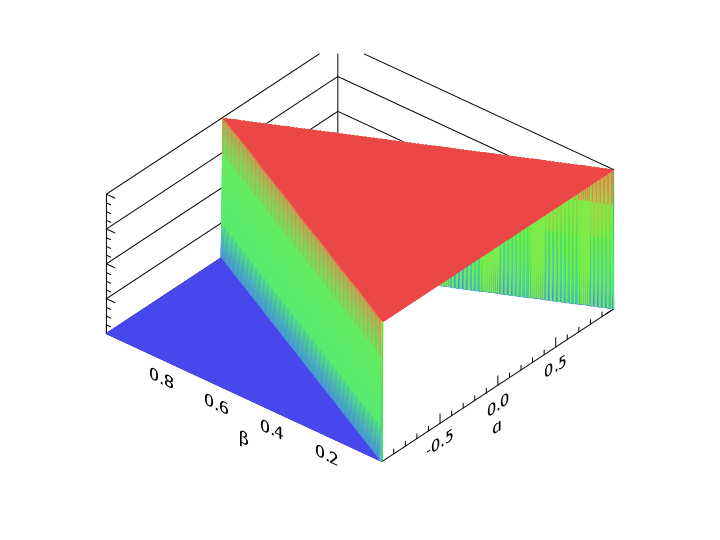}
	}
	\settowidth{\plotwa}{\usebox{\boxplota}}
	
	\sbox{\tableA}{%
		\begin{tabular}{cccccc}

			\toprule

			$n=0$
			\parbox[c]{\plotw}{\includegraphics[scale=0.125]{Plots/R/plots-1}} &
			\parbox[c]{\plotwa}{\includegraphics[scale=0.18]{Plots/R/plots-2}} &
			$n=5$ &
			\parbox[c]{\plotw}{\includegraphics[scale=0.125]{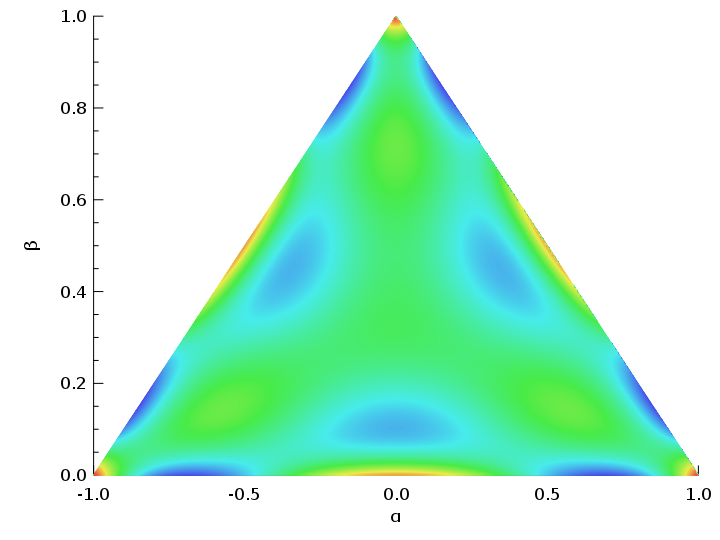}} &
			\parbox[c]{\plotwa}{\includegraphics[scale=0.18]{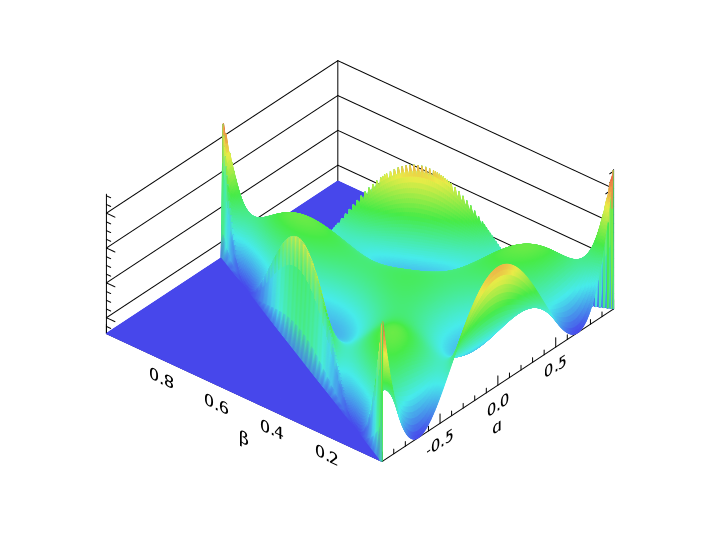}}
			\\

			$n=1$ 
			\parbox[c]{\plotw}{\includegraphics[scale=0.125]{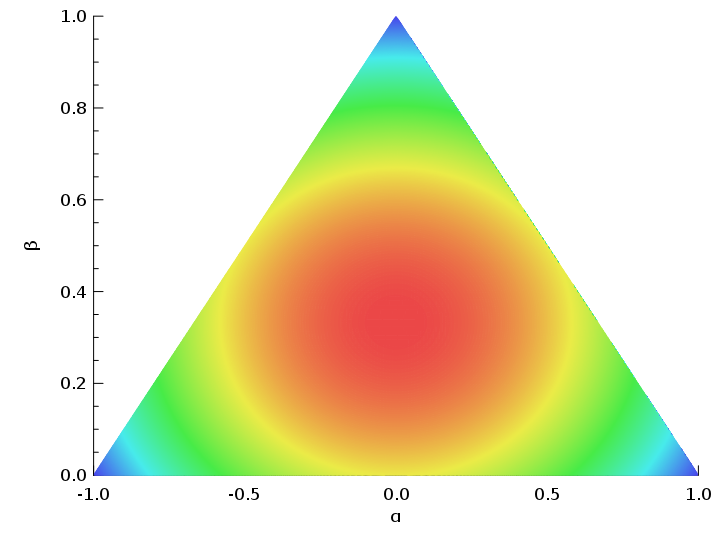}} &
			\parbox[c]{\plotwa}{\includegraphics[scale=0.18]{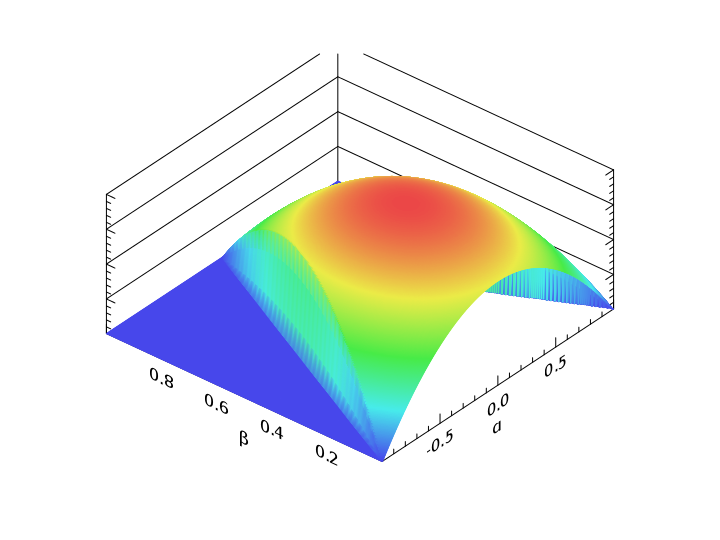}} &
			$n=6$ &
			\parbox[c]{\plotw}{\includegraphics[scale=0.125]{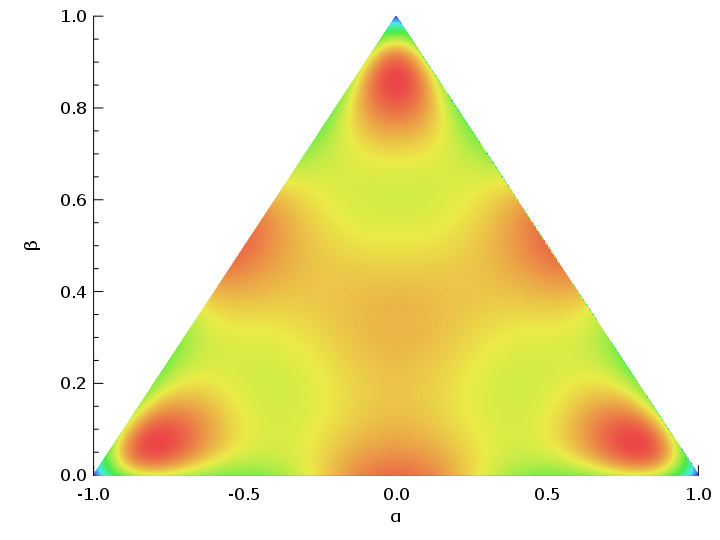}} &
			\parbox[c]{\plotwa}{\includegraphics[scale=0.18]{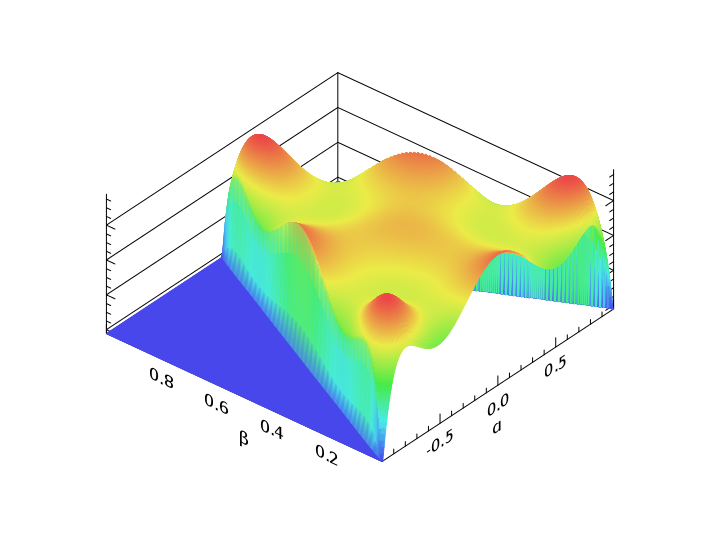}}
			\\

			$n=2$ 
			\parbox[c]{\plotw}{\includegraphics[scale=0.125]{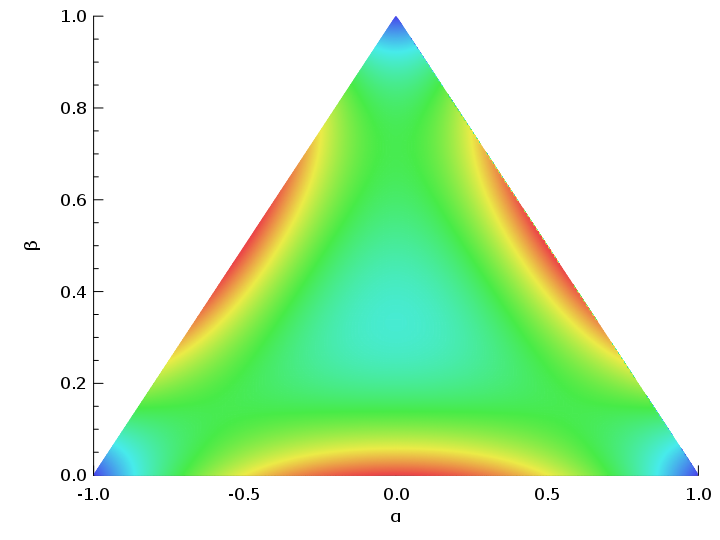}} &
			\parbox[c]{\plotwa}{\includegraphics[scale=0.18]{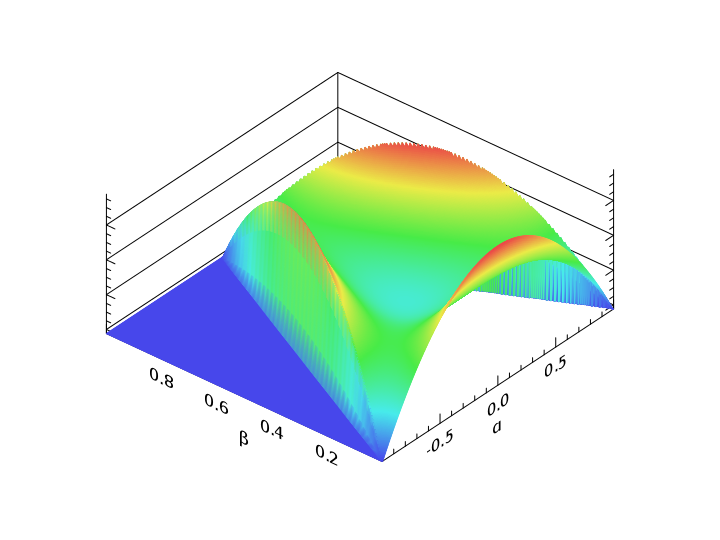}} &
			$n=7$ &
			\parbox[c]{\plotw}{\includegraphics[scale=0.125]{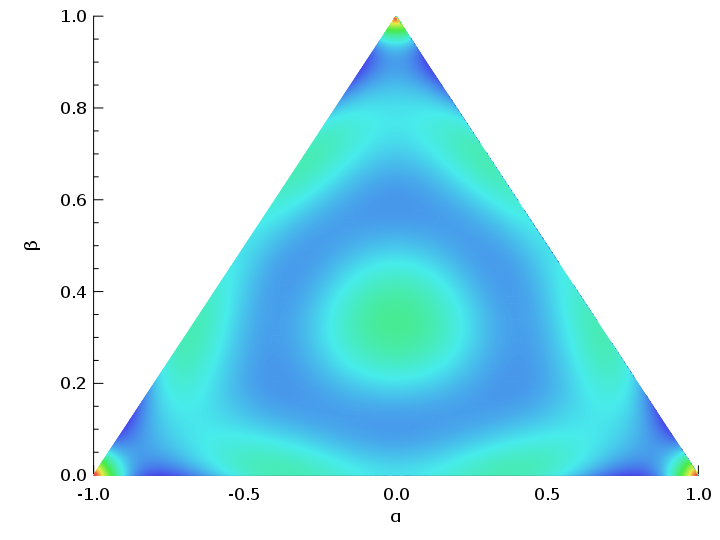}} &
			\parbox[c]{\plotwa}{\includegraphics[scale=0.18]{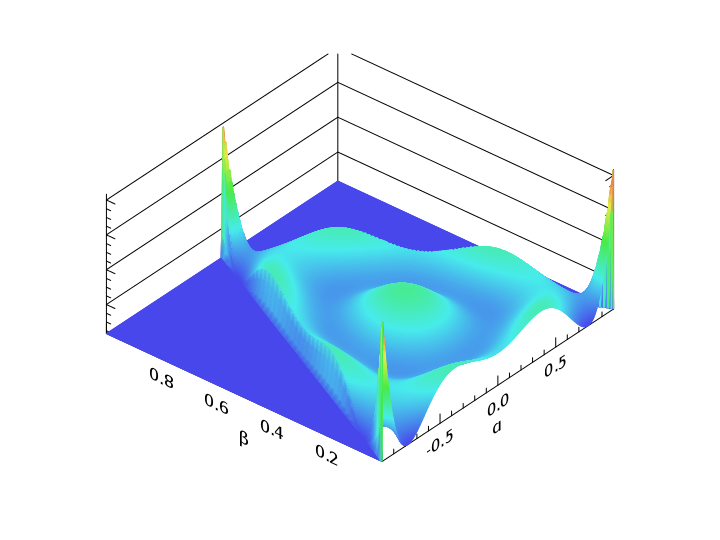}}
			\\

			$n=3$ 
			\parbox[c]{\plotw}{\includegraphics[scale=0.125]{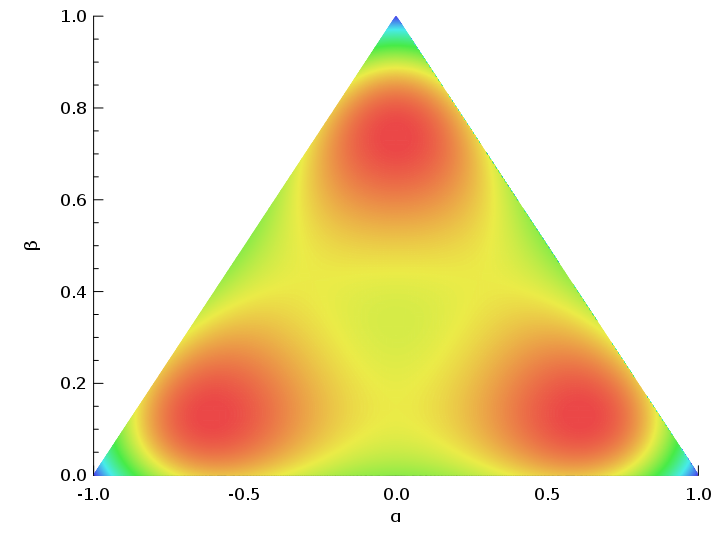}} &
			\parbox[c]{\plotwa}{\includegraphics[scale=0.18]{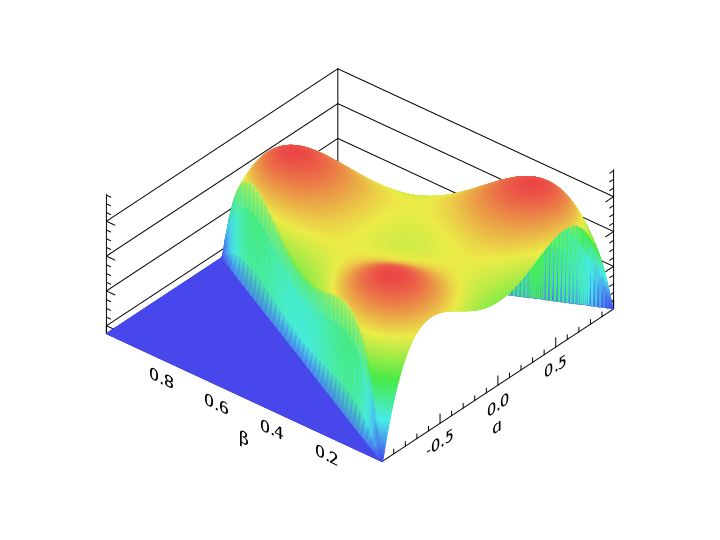}} &
			$n=8$ &
			\parbox[c]{\plotw}{\includegraphics[scale=0.125]{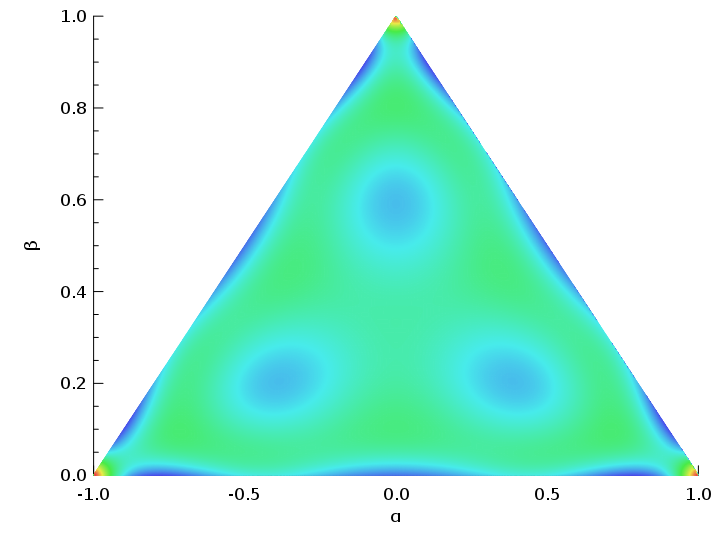}} &
			\parbox[c]{\plotwa}{\includegraphics[scale=0.18]{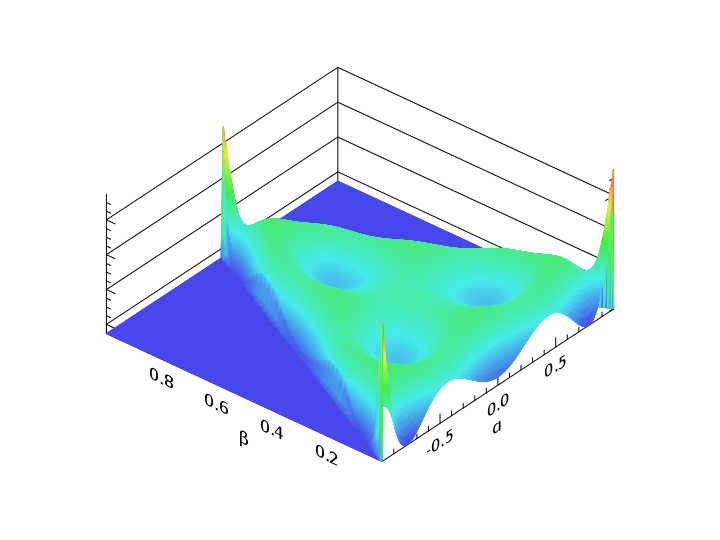}}
			\\

			$n=4$ 
			\parbox[c]{\plotw}{\includegraphics[scale=0.125]{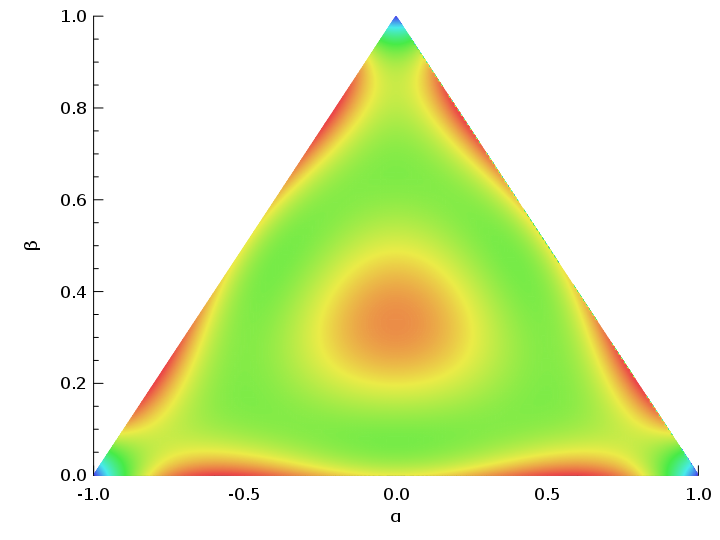}} &
			\parbox[c]{\plotwa}{\includegraphics[scale=0.18]{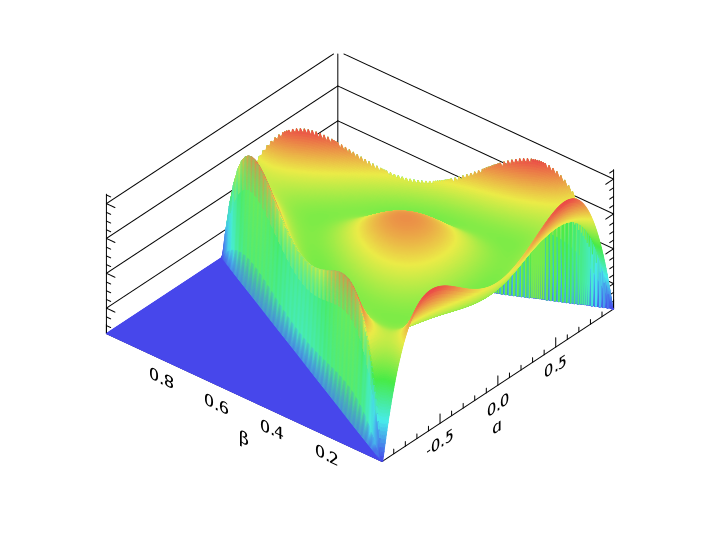}} &
			$n=9$ &
			\parbox[c]{\plotw}{\includegraphics[scale=0.125]{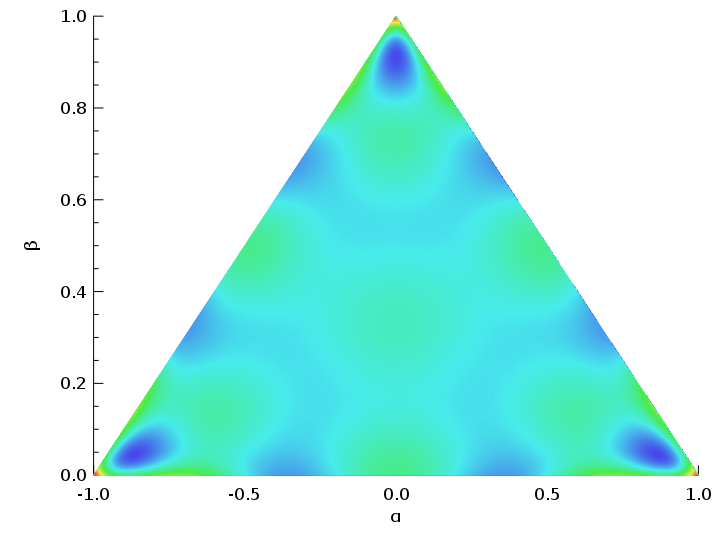}} &
			\parbox[c]{\plotwa}{\includegraphics[scale=0.18]{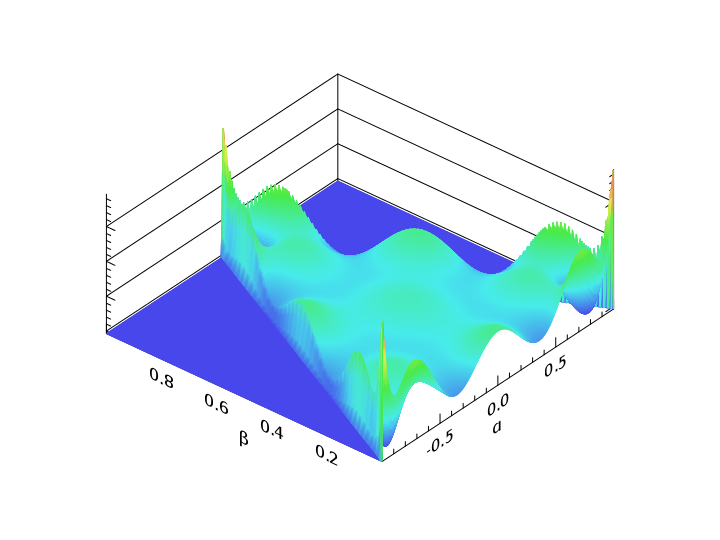}}
			\\

			\bottomrule

		\end{tabular}
	}
	\settowidth{\tblw}{\usebox{\tableA}}
	\addtolength{\tblw}{-1em}

	\begin{center}
		\usebox{\tableA}
	\end{center}
			
	\caption{\label{table:Rbasis}
	Basis shapes $\Rbasis{n}$.}

	\end{sidewaystable}

\begin{table}[h]

	\tablepreamble
		
	\sbox{\tableA}{%
		\begin{tabular}{QqQqQqQqQqQ}

			\toprule

			&
			\multicolumn{1}{c}{$\alpha_0$} &
			\multicolumn{1}{c}{$\alpha_1$} &
			\multicolumn{1}{c}{$\alpha_2$} &
			\multicolumn{1}{c}{$\alpha_3$} &
			\multicolumn{1}{c}{$\alpha_4$} &
			\multicolumn{1}{c}{$\alpha_5$} &
			\multicolumn{1}{c}{$\alpha_6$} &
			\multicolumn{1}{c}{$\alpha_7$} &
			\multicolumn{1}{c}{$\alpha_8$} &
			\multicolumn{1}{c}{$\alpha_9$}

			\\

			\cmidrule{2-11}

			\text{{\Local}} &
			-2.16 & 1.78 & 0.75 & -1.21 & 0.79 &-0.49 & 0.85 & 1.01 &-0.53  & -0.55\\

			\cmidrule{2-11}

			\text{{\Equilateral}} &
			0.52 & 0.23 & -0.16 & -0.03 & -0.01 & 0.02 & 0.00 & 0.05 & 0.02 & -0.01 \\

			\cmidrule{2-11}

			\text{orthogonal} &
			-0.44 & 0.68 & -0.49 & -0.10 & -0.04 & 0.07 & 0.01 & 0.13 & 0.05 & -0.03 \\

			\cmidrule{2-11}

			\text{{\Enfolded}} &
			0.48 & -0.23 & 0.16 & 0.03 & 0.01 & -0.02 & 0.00 &  -0.04  & -0.02 & 0.01\\

			\bottomrule

		\end{tabular}
	}
	\settowidth{\tblw}{\usebox{\tableA}}
	\addtolength{\tblw}{-1em}

	\begin{center}
		\usebox{\tableA}
	\end{center}

	\caption{\label{table:Rtemplates}Expansion of common templates
	in terms of the $\Rbasis{n}$ basis.}
	
	\end{table}

\begin{table}[h]

	\tablepreamble
		
	\sbox{\tableA}{%
		\begin{tabular}{QqQq}
			 	\toprule

			&
			\multicolumn{3}{c}{Approximations to orthogonal shapes}
			\\

			\cmidrule{2-4}

		 	&
		 	\multicolumn{1}{c}{\text{$S_H$ shape}} &
		 	\multicolumn{1}{c}{\text{Creminelli \etal\, shape}\tmark{a}} &
		 	\multicolumn{1}{c}{\text{$P(X,\phi)$ shape}\tmark{b}} 
		 	
		 	\\

			\midrule

			\text{$S_H$ shape} &

			 0.99&
			 0.98 & 0.89
			 
			\\

			\cmidrule{2-4}

			\text{Creminelli \etal \, shape\tmark{a}} &
			 	0.97& 0.99
			  & 0.88

			\\

			\cmidrule{2-4}

			\text{$P(X,\phi)$ shape\tmark{b}} &
				0.82  &
			0.83
			 & 1.00
			
			\\

			\bottomrule

		\end{tabular}
	}
	\settowidth{\tblw}{\usebox{\tableA}}
	\addtolength{\tblw}{-1em}

	\begin{center}
		\usebox{\tableA}
	\end{center}

	\renewcommand{\arraystretch}{1}
	
	\sbox{\tableB}{%
		\begin{tabular}{l@{\hspace{1mm}}l}
			\tmark{a} & \parbox[t]{\tblw}{%
				This is the shape investigated 
				by Creminelli~{\etal}~\cite{Creminelli:2010qf}.}\\
				\tmark{b} & \parbox[t]{\tblw}{%
				This is the shape $O$ constructed from contributions 
				to the bispectrum at next-order in slow-roll 
				\cite{Burrage:2011hd}.}
		\end{tabular}
	}
	
	\begin{center}
		\usebox{\tableB}
	\end{center}

	\caption{\label{table:cosines-approx}Cosines between 
	$\Rbasis{n}$-approximations to the orthogonal shapes
	depicted in table~\ref{table:ortho-approx}
	and the corresponding exact shape.}
	
	\end{table}

	\begin{sidewaystable}[h]

	\tablepreamble
		
	\sbox{\tableA}{%
		\begin{tabular}{QqQqQqQqQqQ}

			\toprule

			&
			\multicolumn{1}{c}{$\alpha_0$} &
			\multicolumn{1}{c}{$\alpha_1$} &
			\multicolumn{1}{c}{$\alpha_2$} &
			\multicolumn{1}{c}{$\alpha_3$} &
			\multicolumn{1}{c}{$\alpha_4$} &
			\multicolumn{1}{c}{$\alpha_5$} &
			\multicolumn{1}{c}{$\alpha_6$} &
			\multicolumn{1}{c}{$\alpha_7$} &
			\multicolumn{1}{c}{$\alpha_8$} &
			\multicolumn{1}{c}{$\alpha_9$}

			\\

			\cmidrule{2-11}

			\text{$S_H$-shape} &
			0.000 & 0.006 & 0.006 & 0.010 & 0.005 & -0.002 & -0.002 & -0.005 &-0.003  & 0.002\\

			\cmidrule{2-11}

			\text{$P(X,\phi)$ shape $O$} &
			0.047 & -0.190 & -0.100 & -0.107 & -0.024 & 0.026 & 0.028 & 0.043 & 0.024 & -0.019 \\

			\cmidrule{2-11}

			\text{Creminelli \etal\, shape} &
			0.000 & 0.015 & 0.019 & 0.024 & 0.011 & -0.006 & -0.002 & -0.012 & -0.006 & 0.003 \\

			\bottomrule

		\end{tabular}
	}
	\settowidth{\tblw}{\usebox{\tableA}}
	\addtolength{\tblw}{-1em}

	\begin{center}
		\usebox{\tableA}
	\end{center}

	\caption{\label{table:Rorthogonal}Expansion of the $S_H$-shape,
	 the $P(X,\phi)$ shape $O$ (at next-order)
	\cite{Burrage:2011hd} and 
	the Creminelli~{\etal} shape \cite{Creminelli:2010qf}
	in terms of the $\Rbasis{n}$ basis.}
	
	\end{sidewaystable}

\clearpage
	\bibliographystyle{JHEPmodplain}
	\bibliography{paper}

\providecommand{\href}[2]{#2}\begingroup\raggedright\begin{thebibliography}{10}

\bibitem{Komatsu:2010fb}
{\bf WMAP Collaboration} Collaboration, E.~Komatsu {\em et~al.}, {\it
  {Seven-Year Wilkinson Microwave Anisotropy Probe (WMAP) Observations:
  Cosmological Interpretation}},  {\sl Astrophys.J.Suppl.} {\bf 192} (2011) 18,
  [\href{http://arxiv.org/abs/1001.4538}{{\sf arXiv:1001.4538}}],
  [\href{http://dx.doi.org/10.1088/0067-0049/192/2/18}{{\sf
  doi:10.1088/0067-0049/192/2/18}}].

\bibitem{Larson:2010gs}
D.~Larson, J.~Dunkley, G.~Hinshaw, E.~Komatsu, M.~Nolta, {\em et~al.}, {\it
  {Seven-Year Wilkinson Microwave Anisotropy Probe (WMAP) Observations: Power
  Spectra and WMAP-Derived Parameters}},  {\sl Astrophys.J.Suppl.} {\bf 192}
  (2011) 16, [\href{http://arxiv.org/abs/1001.4635}{{\sf arXiv:1001.4635}}],
  [\href{http://dx.doi.org/10.1088/0067-0049/192/2/16}{{\sf
  doi:10.1088/0067-0049/192/2/16}}].

\bibitem{Jarosik:2010iu}
N.~Jarosik, C.~Bennett, J.~Dunkley, B.~Gold, M.~Greason, {\em et~al.}, {\it
  {Seven-Year Wilkinson Microwave Anisotropy Probe (WMAP) Observations: Sky
  Maps, Systematic Errors, and Basic Results}},  {\sl Astrophys.J.Suppl.} {\bf
  192} (2011) 14, [\href{http://arxiv.org/abs/1001.4744}{{\sf
  arXiv:1001.4744}}], [\href{http://dx.doi.org/10.1088/0067-0049/192/2/14}{{\sf
  doi:10.1088/0067-0049/192/2/14}}].

\bibitem{Guth:1980zm}
A.~H. Guth, {\it {The Inflationary Universe: A Possible Solution to the Horizon
  and Flatness Problems}},  {\sl Phys.Rev.} {\bf D23} (1981) 347--356,
  [\href{http://dx.doi.org/10.1103/PhysRevD.23.347}{{\sf
  doi:10.1103/PhysRevD.23.347}}].

\bibitem{Guth:2005zr}
A.~H. Guth and D.~I. Kaiser, {\it {Inflationary cosmology: Exploring the
  Universe from the smallest to the largest scales}},  {\sl Science} {\bf 307}
  (2005) 884--890, [\href{http://arxiv.org/abs/astro-ph/0502328}{{\sf
  arXiv:astro-ph/0502328}}],
  [\href{http://dx.doi.org/10.1126/science.1107483}{{\sf
  doi:10.1126/science.1107483}}].

\bibitem{Bardeen:1983qw}
J.~M. Bardeen, P.~J. Steinhardt, and M.~S. Turner, {\it {Spontaneous Creation
  of Almost Scale - Free Density Perturbations in an Inflationary Universe}},
  {\sl Phys.Rev.} {\bf D28} (1983) 679,
  [\href{http://dx.doi.org/10.1103/PhysRevD.28.679}{{\sf
  doi:10.1103/PhysRevD.28.679}}].

\bibitem{Komatsu:2009kd}
E.~Komatsu, N.~Afshordi, N.~Bartolo, D.~Baumann, J.~Bond, {\em et~al.}, {\it
  {Non-Gaussianity as a Probe of the Physics of the Primordial Universe and the
  Astrophysics of the Low Redshift Universe}},
  \href{http://arxiv.org/abs/0902.4759}{{\sf arXiv:0902.4759}}.

\bibitem{Babich:2004gb}
D.~Babich, P.~Creminelli, and M.~Zaldarriaga, {\it {The shape of
  non-Gaussianities}},  {\sl JCAP} {\bf 0408} (2004) 009,
  [\href{http://arxiv.org/abs/astro-ph/0405356}{{\sf arXiv:astro-ph/0405356}}],
  [\href{http://dx.doi.org/10.1088/1475-7516/2004/08/009}{{\sf
  doi:10.1088/1475-7516/2004/08/009}}].

\bibitem{Chen:2010xka}
X.~Chen, {\it {Primordial Non-Gaussianities from Inflation Models}},  {\sl
  Adv.Astron.} {\bf 2010} (2010) 638979,
  [\href{http://arxiv.org/abs/1002.1416}{{\sf arXiv:1002.1416}}],
  [\href{http://dx.doi.org/10.1155/2010/638979}{{\sf
  doi:10.1155/2010/638979}}].

\bibitem{Koyama:2010xj}
K.~Koyama, {\it {Non-Gaussianity of quantum fields during inflation}},  {\sl
  Class.Quant.Grav.} {\bf 27} (2010) 124001,
  [\href{http://arxiv.org/abs/1002.0600}{{\sf arXiv:1002.0600}}],
  [\href{http://dx.doi.org/10.1088/0264-9381/27/12/124001}{{\sf
  doi:10.1088/0264-9381/27/12/124001}}].

\bibitem{Burrage:2010cu}
C.~Burrage, C.~de~Rham, D.~Seery, and A.~J. Tolley, {\it {Galileon inflation}},
   {\sl JCAP} {\bf 1101} (2011) 014,
  [\href{http://arxiv.org/abs/1009.2497}{{\sf arXiv:1009.2497}}],
  [\href{http://dx.doi.org/10.1088/1475-7516/2011/01/014}{{\sf
  doi:10.1088/1475-7516/2011/01/014}}].

\bibitem{Baumann:2011su}
D.~Baumann and D.~Green, {\it {Equilateral Non-Gaussianity and New Physics on
  the Horizon}},  {\sl JCAP} {\bf 1109} (2011) 014,
  [\href{http://arxiv.org/abs/1102.5343}{{\sf arXiv:1102.5343}}],
  [\href{http://dx.doi.org/10.1088/1475-7516/2011/09/014}{{\sf
  doi:10.1088/1475-7516/2011/09/014}}]. * Temporary entry *.

\bibitem{Creminelli:2010qf}
P.~Creminelli, G.~D'Amico, M.~Musso, J.~Norena, and E.~Trincherini, {\it
  {Galilean symmetry in the effective theory of inflation: new shapes of
  non-Gaussianity}},  {\sl JCAP} {\bf 1102} (2011) 006,
  [\href{http://arxiv.org/abs/1011.3004}{{\sf arXiv:1011.3004}}],
  [\href{http://dx.doi.org/10.1088/1475-7516/2011/02/006}{{\sf
  doi:10.1088/1475-7516/2011/02/006}}].

\bibitem{Gao:2011qe}
X.~Gao and D.~A. Steer, {\it {Inflation and primordial non-Gaussianities of
  ``generalized Galileons''}},  \href{http://arxiv.org/abs/1107.2642}{{\sf
  arXiv:1107.2642}}.

\bibitem{DeFelice:2011uc}
A.~De~Felice and S.~Tsujikawa, {\it {Inflationary non-Gaussianities in the most
  general second-order scalar-tensor theories}},
  \href{http://arxiv.org/abs/1107.3917}{{\sf arXiv:1107.3917}}.

\bibitem{RenauxPetel:2011sb}
S.~Renaux-Petel, {\it {On the redundancy of operators and the bispectrum in the
  most general second-order scalar-tensor theory}},
  \href{http://arxiv.org/abs/1107.5020}{{\sf arXiv:1107.5020}}.

\bibitem{RenauxPetel:2011uk}
S.~Renaux-Petel, S.~Mizuno, and K.~Koyama, {\it {Primordial fluctuations and
  non-Gaussianities from multifield DBI Galileon inflation}},
  \href{http://arxiv.org/abs/1108.0305}{{\sf arXiv:1108.0305}}.

\bibitem{Vazquez:2008wb}
S.~E. Vazquez, {\it {Constraining Modified Gravity with Large
  non-Gaussianities}},  {\sl Phys.Rev.} {\bf D79} (2009) 043520,
  [\href{http://arxiv.org/abs/0806.0603}{{\sf arXiv:0806.0603}}],
  [\href{http://dx.doi.org/10.1103/PhysRevD.79.043520}{{\sf
  doi:10.1103/PhysRevD.79.043520}}].

\bibitem{Gao:2010um}
X.~Gao, {\it {Testing gravity with non-Gaussianity}},  {\sl Phys.Lett.} {\bf
  B702} (2011) 197--200, [\href{http://arxiv.org/abs/1008.2123}{{\sf
  arXiv:1008.2123}}],
  [\href{http://dx.doi.org/10.1016/j.physletb.2011.07.022}{{\sf
  doi:10.1016/j.physletb.2011.07.022}}].

\bibitem{deRham:2010ik}
C.~de~Rham and G.~Gabadadze, {\it {Generalization of the Fierz-Pauli Action}},
  {\sl Phys.Rev.} {\bf D82} (2010) 044020,
  [\href{http://arxiv.org/abs/1007.0443}{{\sf arXiv:1007.0443}}],
  [\href{http://dx.doi.org/10.1103/PhysRevD.82.044020}{{\sf
  doi:10.1103/PhysRevD.82.044020}}].

\bibitem{deRham:2010kj}
C.~de~Rham, G.~Gabadadze, and A.~J. Tolley, {\it {Resummation of Massive
  Gravity}},  {\sl Phys.Rev.Lett.} {\bf 106} (2011) 231101,
  [\href{http://arxiv.org/abs/1011.1232}{{\sf arXiv:1011.1232}}],
  [\href{http://dx.doi.org/10.1103/PhysRevLett.106.231101}{{\sf
  doi:10.1103/PhysRevLett.106.231101}}].

\bibitem{Hassan:2011hr}
S.~Hassan and R.~A. Rosen, {\it {Resolving the Ghost Problem in non-Linear
  Massive Gravity}},  \href{http://arxiv.org/abs/1106.3344}{{\sf
  arXiv:1106.3344}}.

\bibitem{Hinterbichler:2011tt}
K.~Hinterbichler, {\it {Theoretical Aspects of Massive Gravity}},
  \href{http://arxiv.org/abs/1105.3735}{{\sf arXiv:1105.3735}}.

\bibitem{Dvali:2000hr}
G.~Dvali, G.~Gabadadze, and M.~Porrati, {\it {4-D gravity on a brane in 5-D
  Minkowski space}},  {\sl Phys.Lett.} {\bf B485} (2000) 208--214,
  [\href{http://arxiv.org/abs/hep-th/0005016}{{\sf arXiv:hep-th/0005016}}],
  [\href{http://dx.doi.org/10.1016/S0370-2693(00)00669-9}{{\sf
  doi:10.1016/S0370-2693(00)00669-9}}].

\bibitem{Deffayet:2001pu}
C.~Deffayet, G.~Dvali, and G.~Gabadadze, {\it {Accelerated universe from
  gravity leaking to extra dimensions}},  {\sl Phys.Rev.} {\bf D65} (2002)
  044023, [\href{http://arxiv.org/abs/astro-ph/0105068}{{\sf
  arXiv:astro-ph/0105068}}],
  [\href{http://dx.doi.org/10.1103/PhysRevD.65.044023}{{\sf
  doi:10.1103/PhysRevD.65.044023}}].

\bibitem{Kobayashi:2010cm}
T.~Kobayashi, M.~Yamaguchi, and J.~Yokoyama, {\it {G-inflation: Inflation
  driven by the Galileon field}},  {\sl Phys.Rev.Lett.} {\bf 105} (2010)
  231302, [\href{http://arxiv.org/abs/1008.0603}{{\sf arXiv:1008.0603}}],
  [\href{http://dx.doi.org/10.1103/PhysRevLett.105.231302}{{\sf
  doi:10.1103/PhysRevLett.105.231302}}].

\bibitem{Mizuno:2010ag}
S.~Mizuno and K.~Koyama, {\it {Primordial non-Gaussianity from the DBI
  Galileons}},  {\sl Phys.Rev.} {\bf D82} (2010) 103518,
  [\href{http://arxiv.org/abs/1009.0677}{{\sf arXiv:1009.0677}}],
  [\href{http://dx.doi.org/10.1103/PhysRevD.82.103518}{{\sf
  doi:10.1103/PhysRevD.82.103518}}].

\bibitem{Kobayashi:2011pc}
T.~Kobayashi, M.~Yamaguchi, and J.~Yokoyama, {\it {Primordial non-Gaussianity
  from G-inflation}},  {\sl Phys.Rev.D} (2011)
  [\href{http://arxiv.org/abs/1103.1740}{{\sf arXiv:1103.1740}}].

\bibitem{RenauxPetel:2011dv}
S.~Renaux-Petel, {\it {Orthogonal non-Gaussianities from Dirac-Born-Infeld
  Galileon inflation}},  {\sl Class.Quant.Grav.} {\bf 28} (2011) 182001,
  [\href{http://arxiv.org/abs/1105.6366}{{\sf arXiv:1105.6366}}],
  [\href{http://dx.doi.org/10.1088/0264-9381/28/18/182001}{{\sf
  doi:10.1088/0264-9381/28/18/182001}}]. * Temporary entry *.

\bibitem{Kobayashi:2011nu}
T.~Kobayashi, M.~Yamaguchi, and J.~Yokoyama, {\it {Generalized G-inflation:
  Inflation with the most general second-order field equations}},
  \href{http://arxiv.org/abs/1105.5723}{{\sf arXiv:1105.5723}}.

\bibitem{Horndeski:1974}
G.~W. Horndeski {\sl Int. J. Theor. Phys.} {\bf 10} (1974) 363.

\bibitem{Charmousis:2011bf}
C.~Charmousis, E.~J. Copeland, A.~Padilla, and P.~M. Saffin, {\it {General
  second order scalar-tensor theory, self tuning, and the Fab Four}},
  \href{http://arxiv.org/abs/1106.2000}{{\sf arXiv:1106.2000}}.

\bibitem{Deffayet:2009mn}
C.~Deffayet, S.~Deser, and G.~Esposito-Farese, {\it {Generalized Galileons: All
  scalar models whose curved background extensions maintain second-order field
  equations and stress-tensors}},  {\sl Phys.Rev.} {\bf D80} (2009) 064015,
  [\href{http://arxiv.org/abs/0906.1967}{{\sf arXiv:0906.1967}}],
  [\href{http://dx.doi.org/10.1103/PhysRevD.80.064015}{{\sf
  doi:10.1103/PhysRevD.80.064015}}].

\bibitem{Deffayet:2011gz}
C.~Deffayet, X.~Gao, D.~A. Steer, and G.~Zahariade, {\it {From k-essence to
  generalised Galileons}},  \href{http://arxiv.org/abs/1103.3260}{{\sf
  arXiv:1103.3260}}.

\bibitem{Cheung:2007st}
C.~Cheung, P.~Creminelli, A.~Fitzpatrick, J.~Kaplan, and L.~Senatore, {\it {The
  Effective Field Theory of Inflation}},  {\sl JHEP} {\bf 0803} (2008) 014,
  [\href{http://arxiv.org/abs/0709.0293}{{\sf arXiv:0709.0293}}],
  [\href{http://dx.doi.org/10.1088/1126-6708/2008/03/014}{{\sf
  doi:10.1088/1126-6708/2008/03/014}}].

\bibitem{Burrage:2011hd}
C.~Burrage, R.~H. Ribeiro, and D.~Seery, {\it {Large slow-roll corrections to
  the bispectrum of noncanonical inflation}},  {\sl JCAP} {\bf 1107} (2011)
  032, [\href{http://arxiv.org/abs/1103.4126}{{\sf arXiv:1103.4126}}],
  [\href{http://dx.doi.org/10.1088/1475-7516/2011/07/032}{{\sf
  doi:10.1088/1475-7516/2011/07/032}}]. * Temporary entry *.

\bibitem{Fergusson:2009nv}
J.~Fergusson, M.~Liguori, and E.~Shellard, {\it {General CMB and Primordial
  Bispectrum Estimation I: Mode Expansion, Map-Making and Measures of
  {$\fNL$}}},  {\sl Phys.Rev.} {\bf D82} (2010) 023502,
  [\href{http://arxiv.org/abs/0912.5516}{{\sf arXiv:0912.5516}}],
  [\href{http://dx.doi.org/10.1103/PhysRevD.82.023502}{{\sf
  doi:10.1103/PhysRevD.82.023502}}].

\bibitem{Nicolis:2008in}
A.~Nicolis, R.~Rattazzi, and E.~Trincherini, {\it {The Galileon as a local
  modification of gravity}},  {\sl Phys.Rev.} {\bf D79} (2009) 064036,
  [\href{http://arxiv.org/abs/0811.2197}{{\sf arXiv:0811.2197}}],
  [\href{http://dx.doi.org/10.1103/PhysRevD.79.064036}{{\sf
  doi:10.1103/PhysRevD.79.064036}}].

\bibitem{deRham:2010eu}
C.~de~Rham and A.~J. Tolley, {\it {DBI and the Galileon reunited}},  {\sl JCAP}
  {\bf 1005} (2010) 015, [\href{http://arxiv.org/abs/1003.5917}{{\sf
  arXiv:1003.5917}}],
  [\href{http://dx.doi.org/10.1088/1475-7516/2010/05/015}{{\sf
  doi:10.1088/1475-7516/2010/05/015}}].

\bibitem{Deffayet:2009wt}
C.~Deffayet, G.~Esposito-Farese, and A.~Vikman, {\it {Covariant Galileon}},
  {\sl Phys.Rev.} {\bf D79} (2009) 084003,
  [\href{http://arxiv.org/abs/0901.1314}{{\sf arXiv:0901.1314}}],
  [\href{http://dx.doi.org/10.1103/PhysRevD.79.084003}{{\sf
  doi:10.1103/PhysRevD.79.084003}}].

\bibitem{Goon:2011qf}
G.~Goon, K.~Hinterbichler, and M.~Trodden, {\it {Symmetries for Galileons and
  DBI scalars on curved space}},  {\sl JCAP} {\bf 1107} (2011) 017,
  [\href{http://arxiv.org/abs/1103.5745}{{\sf arXiv:1103.5745}}],
  [\href{http://dx.doi.org/10.1088/1475-7516/2011/07/017}{{\sf
  doi:10.1088/1475-7516/2011/07/017}}].

\bibitem{Goon:2011uw}
G.~Goon, K.~Hinterbichler, and M.~Trodden, {\it {A New Class of Effective Field
  Theories from Embedded Branes}},  {\sl Phys.Rev.Lett.} {\bf 106} (2011)
  231102, [\href{http://arxiv.org/abs/1103.6029}{{\sf arXiv:1103.6029}}],
  [\href{http://dx.doi.org/10.1103/PhysRevLett.106.231102}{{\sf
  doi:10.1103/PhysRevLett.106.231102}}].

\bibitem{Trodden:2011xh}
M.~Trodden and K.~Hinterbichler, {\it {Generalizing Galileons}},  {\sl
  Class.Quant.Grav.} {\bf 28} (2011) 204003,
  [\href{http://arxiv.org/abs/1104.2088}{{\sf arXiv:1104.2088}}],
  [\href{http://dx.doi.org/10.1088/0264-9381/28/20/204003}{{\sf
  doi:10.1088/0264-9381/28/20/204003}}].

\bibitem{Burrage:2011bt}
C.~Burrage, C.~de~Rham, and L.~Heisenberg, {\it {de Sitter Galileon}},  {\sl
  JCAP} {\bf 1105} (2011) 025, [\href{http://arxiv.org/abs/1104.0155}{{\sf
  arXiv:1104.0155}}],
  [\href{http://dx.doi.org/10.1088/1475-7516/2011/05/025}{{\sf
  doi:10.1088/1475-7516/2011/05/025}}].

\bibitem{Vainshtein:1972sx}
A.~Vainshtein, {\it {To the problem of nonvanishing gravitation mass}},  {\sl
  Phys.Lett.} {\bf B39} (1972) 393--394,
  [\href{http://dx.doi.org/10.1016/0370-2693(72)90147-5}{{\sf
  doi:10.1016/0370-2693(72)90147-5}}].

\bibitem{Burrage:2010rs}
C.~Burrage and D.~Seery, {\it {Revisiting fifth forces in the Galileon model}},
   {\sl JCAP} {\bf 1008} (2010) 011,
  [\href{http://arxiv.org/abs/1005.1927}{{\sf arXiv:1005.1927}}],
  [\href{http://dx.doi.org/10.1088/1475-7516/2010/08/011}{{\sf
  doi:10.1088/1475-7516/2010/08/011}}].

\bibitem{Gannouji:2010au}
R.~Gannouji and M.~Sami, {\it {Galileon gravity and its relevance to late time
  cosmic acceleration}},  {\sl Phys. Rev.} {\bf D82} (2010) 024011,
  [\href{http://arxiv.org/abs/1004.2808}{{\sf arXiv:1004.2808}}],
  [\href{http://dx.doi.org/10.1103/PhysRevD.82.024011}{{\sf
  doi:10.1103/PhysRevD.82.024011}}].

\bibitem{Ali:2010gr}
A.~Ali, R.~Gannouji, and M.~Sami, {\it {Modified gravity a la Galileon: Late
  time cosmic acceleration and observational constraints}},  {\sl Phys.Rev.}
  {\bf D82} (2010) 103015, [\href{http://arxiv.org/abs/1008.1588}{{\sf
  arXiv:1008.1588}}], [\href{http://dx.doi.org/10.1103/PhysRevD.82.103015}{{\sf
  doi:10.1103/PhysRevD.82.103015}}].

\bibitem{Brax:2011sv}
P.~Brax, C.~Burrage, and A.-C. Davis, {\it {Laboratory Tests of the Galileon}},
   {\sl JCAP} {\bf 1109} (2011) 020,
  [\href{http://arxiv.org/abs/1106.1573}{{\sf arXiv:1106.1573}}],
  [\href{http://dx.doi.org/10.1088/1475-7516/2011/09/020}{{\sf
  doi:10.1088/1475-7516/2011/09/020}}].

\bibitem{Deffayet:2010qz}
C.~Deffayet, O.~Pujolas, I.~Sawicki, and A.~Vikman, {\it {Imperfect Dark Energy
  from Kinetic Gravity Braiding}},  {\sl JCAP} {\bf 1010} (2010) 026,
  [\href{http://arxiv.org/abs/1008.0048}{{\sf arXiv:1008.0048}}],
  [\href{http://dx.doi.org/10.1088/1475-7516/2010/10/026}{{\sf
  doi:10.1088/1475-7516/2010/10/026}}].

\bibitem{Fergusson:2008ra}
J.~R. Fergusson and E.~P.~S. Shellard, {\it {The shape of primordial
  non-Gaussianity and the CMB bispectrum}},  {\sl Phys. Rev.} {\bf D80} (2009)
  043510, [\href{http://arxiv.org/abs/0812.3413}{{\sf arXiv:0812.3413}}],
  [\href{http://dx.doi.org/10.1103/PhysRevD.80.043510}{{\sf
  doi:10.1103/PhysRevD.80.043510}}].

\bibitem{Senatore:2009gt}
L.~Senatore, K.~M. Smith, and M.~Zaldarriaga, {\it {Non-Gaussianities in Single
  Field Inflation and their Optimal Limits from the WMAP 5-year Data}},  {\sl
  JCAP} {\bf 1001} (2010) 028, [\href{http://arxiv.org/abs/0905.3746}{{\sf
  arXiv:0905.3746}}],
  [\href{http://dx.doi.org/10.1088/1475-7516/2010/01/028}{{\sf
  doi:10.1088/1475-7516/2010/01/028}}].

\bibitem{Creminelli:2005hu}
P.~Creminelli, A.~Nicolis, L.~Senatore, M.~Tegmark, and M.~Zaldarriaga, {\it
  {Limits on non-gaussianities from {WMAP} data}},  {\sl JCAP} {\bf 0605}
  (2006) 004, [\href{http://arxiv.org/abs/astro-ph/0509029}{{\sf
  arXiv:astro-ph/0509029}}],
  [\href{http://dx.doi.org/10.1088/1475-7516/2006/05/004}{{\sf
  doi:10.1088/1475-7516/2006/05/004}}].

\bibitem{Creminelli:2006rz}
P.~Creminelli, L.~Senatore, M.~Zaldarriaga, and M.~Tegmark, {\it {Limits on
  $\fNL$ parameters from WMAP 3yr data}},  {\sl JCAP} {\bf 0703} (2007) 005,
  [\href{http://arxiv.org/abs/astro-ph/0610600}{{\sf arXiv:astro-ph/0610600}}],
  [\href{http://dx.doi.org/10.1088/1475-7516/2007/03/005}{{\sf
  doi:10.1088/1475-7516/2007/03/005}}].

\bibitem{Meerburg:2010ks}
P.~Meerburg, {\it {Oscillations in the Primordial Bispectrum I: Mode
  Expansion}},  {\sl Phys.Rev.} {\bf D82} (2010) 063517,
  [\href{http://arxiv.org/abs/1006.2771}{{\sf arXiv:1006.2771}}],
  [\href{http://dx.doi.org/10.1103/PhysRevD.82.063517}{{\sf
  doi:10.1103/PhysRevD.82.063517}}].

\bibitem{Copeland:1993jj}
E.~J. Copeland, E.~W. Kolb, A.~R. Liddle, and J.~E. Lidsey, {\it
  {Reconstructing the inflation potential, in principle and in practice}},
  {\sl Phys.Rev.} {\bf D48} (1993) 2529--2547,
  [\href{http://arxiv.org/abs/hep-ph/9303288}{{\sf arXiv:hep-ph/9303288}}],
  [\href{http://dx.doi.org/10.1103/PhysRevD.48.2529}{{\sf
  doi:10.1103/PhysRevD.48.2529}}].

\bibitem{Copeland:1993zn}
E.~J. Copeland, E.~W. Kolb, A.~R. Liddle, and J.~E. Lidsey, {\it
  {Reconstructing the inflaton potential: Perturbative reconstruction to second
  order}},  {\sl Phys.Rev.} {\bf D49} (1994) 1840--1844,
  [\href{http://arxiv.org/abs/astro-ph/9308044}{{\sf arXiv:astro-ph/9308044}}],
  [\href{http://dx.doi.org/10.1103/PhysRevD.49.1840}{{\sf
  doi:10.1103/PhysRevD.49.1840}}].

\bibitem{Alishahiha:2004eh}
M.~Alishahiha, E.~Silverstein, and D.~Tong, {\it {DBI in the sky}},  {\sl Phys.
  Rev.} {\bf D70} (2004) 123505,
  [\href{http://arxiv.org/abs/hep-th/0404084}{{\sf arXiv:hep-th/0404084}}],
  [\href{http://dx.doi.org/10.1103/PhysRevD.70.123505}{{\sf
  doi:10.1103/PhysRevD.70.123505}}].

\bibitem{Chen:2006nt}
X.~Chen, M.-x. Huang, S.~Kachru, and G.~Shiu, {\it {Observational signatures
  and non-Gaussianities of general single field inflation}},  {\sl JCAP} {\bf
  0701} (2007) 002, [\href{http://arxiv.org/abs/hep-th/0605045}{{\sf
  arXiv:hep-th/0605045}}],
  [\href{http://dx.doi.org/10.1088/1475-7516/2007/01/002}{{\sf
  doi:10.1088/1475-7516/2007/01/002}}].

\bibitem{Franche:2009gk}
P.~Franche, R.~Gwyn, B.~Underwood, and A.~Wissanji, {\it {Attractive
  Lagrangians for Non-Canonical Inflation}},  {\sl Phys.Rev.} {\bf D81} (2010)
  123526, [\href{http://arxiv.org/abs/0912.1857}{{\sf arXiv:0912.1857}}],
  [\href{http://dx.doi.org/10.1103/PhysRevD.81.123526}{{\sf
  doi:10.1103/PhysRevD.81.123526}}].

\bibitem{Lidsey:1995np}
J.~E. Lidsey, A.~R. Liddle, E.~W. Kolb, E.~J. Copeland, T.~Barreiro, {\em
  et~al.}, {\it {Reconstructing the inflation potential : An overview}},  {\sl
  Rev.Mod.Phys.} {\bf 69} (1997) 373--410,
  [\href{http://arxiv.org/abs/astro-ph/9508078}{{\sf arXiv:astro-ph/9508078}}],
  [\href{http://dx.doi.org/10.1103/RevModPhys.69.373}{{\sf
  doi:10.1103/RevModPhys.69.373}}].

\bibitem{ArmendarizPicon:1999rj}
C.~Armendariz-Picon, T.~Damour, and V.~F. Mukhanov, {\it {{$k$}-inflation}},
  {\sl Phys.Lett.} {\bf B458} (1999) 209--218,
  [\href{http://arxiv.org/abs/hep-th/9904075}{{\sf arXiv:hep-th/9904075}}],
  [\href{http://dx.doi.org/10.1016/S0370-2693(99)00603-6}{{\sf
  doi:10.1016/S0370-2693(99)00603-6}}].

\bibitem{Smidt:2010ra}
J.~Smidt, A.~Amblard, C.~T. Byrnes, A.~Cooray, A.~Heavens, {\em et~al.}, {\it
  {CMB Constraints on Primordial non-Gaussianity from the Bispectrum ($\fNL$)
  and Trispectrum ($\gNL$ and $\tauNL$) and a New Consistency Test of
  Single-Field Inflation}},  {\sl Phys.Rev.} {\bf D81} (2010) 123007,
  [\href{http://arxiv.org/abs/1004.1409}{{\sf arXiv:1004.1409}}],
  [\href{http://dx.doi.org/10.1103/PhysRevD.81.123007}{{\sf
  doi:10.1103/PhysRevD.81.123007}}].

\bibitem{Sasaki:2006kq}
M.~Sasaki, J.~Valiviita, and D.~Wands, {\it {Non-Gaussianity of the primordial
  perturbation in the curvaton model}},  {\sl Phys.Rev.} {\bf D74} (2006)
  103003, [\href{http://arxiv.org/abs/astro-ph/0607627}{{\sf
  arXiv:astro-ph/0607627}}],
  [\href{http://dx.doi.org/10.1103/PhysRevD.74.103003}{{\sf
  doi:10.1103/PhysRevD.74.103003}}].

\bibitem{Byrnes:2006vq}
C.~T. Byrnes, M.~Sasaki, and D.~Wands, {\it {The primordial trispectrum from
  inflation}},  {\sl Phys.Rev.} {\bf D74} (2006) 123519,
  [\href{http://arxiv.org/abs/astro-ph/0611075}{{\sf arXiv:astro-ph/0611075}}],
  [\href{http://dx.doi.org/10.1103/PhysRevD.74.123519}{{\sf
  doi:10.1103/PhysRevD.74.123519}}].

\bibitem{Seery:2006js}
D.~Seery and J.~E. Lidsey, {\it {Non-Gaussianity from the inflationary
  trispectrum}},  {\sl JCAP} {\bf 0701} (2007) 008,
  [\href{http://arxiv.org/abs/astro-ph/0611034}{{\sf arXiv:astro-ph/0611034}}],
  [\href{http://dx.doi.org/10.1088/1475-7516/2007/01/008}{{\sf
  doi:10.1088/1475-7516/2007/01/008}}].

\bibitem{Suyama:2007bg}
T.~Suyama and M.~Yamaguchi, {\it {Non-Gaussianity in the modulated reheating
  scenario}},  {\sl PHRVA,D77,023505.2008} {\bf D77} (2008) 023505,
  [\href{http://arxiv.org/abs/0709.2545}{{\sf arXiv:0709.2545}}],
  [\href{http://dx.doi.org/10.1103/PhysRevD.77.023505}{{\sf
  doi:10.1103/PhysRevD.77.023505}}].

\bibitem{Smith:2011if}
K.~M. Smith, M.~LoVerde, and M.~Zaldarriaga, {\it {A universal bound on
  {$N$}-point correlations from inflation}},
  \href{http://arxiv.org/abs/1108.1805}{{\sf arXiv:1108.1805}}.

\end{thebibliography}\endgroup


\providecommand{\href}[2]{#2}\begingroup\raggedright\endgroup

\end{document}